\documentclass[sigplan,screen]{acmart}
\usepackage{graphicx} 
\usepackage{lipsum}
\usepackage[normalem]{ulem}
\usepackage{amsmath}
\usepackage{mathtools}
\usepackage{multirow}
\usepackage{tabularx}
\usepackage{subfigure}
\usepackage{subcaption}
\usepackage{enumitem}
\usepackage[]{hyperref}

\copyrightyear{2024}
\acmYear{2024}
\setcopyright{rightsretained} 
\acmConference[ASPLOS '24]{29th ACM International Conference on Architectural Support for Programming Languages and Operating Systems, Volume 1}{April 27-May 1, 2024}{La Jolla, CA, USA}
\acmBooktitle{29th ACM International Conference on Architectural Support for Programming Languages and Operating Systems, Volume 1 (ASPLOS '24), April 27-May 1, 2024, La Jolla, CA, USA}
\acmDOI{10.1145/3622781.3674179}
\acmISBN{979-8-4007-0391-1/24/04}

\title{\OurScheme{}: Evaluating Large Quantum Circuits on Small Quantum Computers through Integrated Qubit Reuse and Circuit Cutting}

\author{Aditya Pawar}
\email{adp110@pitt.edu}
\affiliation{%
  \institution{University of Pittsburgh}
  \country{USA}
}

\author{Yingheng Li}
\email{yil392@pitt.edu}
\affiliation{%
  \institution{University of Pittsburgh}
  \country{USA}
}

\author{Zewei Mo}
\email{zewei.mo@pitt.edu}
\affiliation{%
  \institution{University of Pittsburgh}
  \country{USA}
}

\author{Yanan Guo}
\email{yag45@pitt.edu}
\affiliation{%
  \institution{University of Pittsburgh}
  \country{USA}
}

\author{Xulong Tang}
\email{tax6@pitt.edu}
\affiliation{%
  \institution{University of Pittsburgh}
  \country{USA}
}

\author{Youtao Zhang}
\email{zhangyt@cs.pitt.edu}
\affiliation{%
  \institution{University of Pittsburgh}
  \country{USA}
}

\author{Jun Yang}
\email{juy9@pitt.edu}
\affiliation{%
  \institution{University of Pittsburgh}
  \country{USA}
}

\date{March 2024}

\renewcommand{\shortauthors}{Aditya Pawar, Yingheng Li, Zewei Mo, Yanan Guo, Xulong Tang, Youtao Zhang, Jun Yang}
\renewcommand{\shorttitle}{QRCC: Evaluating Large Quantum Circuits on Small
Quantum Computers\\ through Integrated Qubit Reuse and Circuit Cutting}
\newcommand{\OurScheme}{QRCC}
\begin{document}

\begin{abstract}
Quantum computing has recently emerged as a promising computing paradigm for many application domains. However, the size of quantum circuits that can be run with high fidelity is constrained by the limited quantity and quality of physical qubits. 
Recently proposed schemes, such as wire cutting and qubit reuse, mitigate the problem but produce sub-optimal results as they address the problem individually. In addition, gate cutting, an alternative circuit-cutting strategy that is suitable for circuits computing expectation values, has not been fully explored in the field.

In this paper, we propose \OurScheme{}, an integrated approach that exploits qubit reuse and circuit-cutting (including wire cutting and gate cutting) to run large circuits on small quantum computers. 
Circuit-cutting techniques introduce non-negligible post-processing overhead, which increases exponentially with the number of cuts. \OurScheme{} exploits qubit reuse to find better cutting solutions to minimize the cut numbers and thus the post-processing overhead.
Our evaluation results show that on average we reduce the number of cuts by 29\% and an additional reduction when considering gate cuts.
\end{abstract}
    
\maketitle 

\section{Introduction} \label{introduction}

Quantum computing has recently emerged as a promising computing paradigm for many application domains, such as machine learning \cite{QNN,ML,ML2}, chemistry simulation \cite{chemistry,chem1,chem2}, and optimization \cite{optimization,finance_optimization}. 
The problems from these domains scale quickly such that they require increasingly larger fault-tolerant quantum computers. Unfortunately, we are currently in the NISQ (noisy intermediate-scale quantum) era \cite{Book} where quantum devices suffer from various noises, e.g., short coherence time and crosstalk among qubits, and small device sizes, e.g., current quantum devices only have up to 100s of qubits.


It has become one of the major challenges to run large quantum circuits in the NISQ era. Large quantum computers, e.g., IBM 433 osprey, often have limited availability to the general public. In addition, not all qubits of large quantum computers exhibit high computational fidelity --- some noisy qubits have to be {\em frozen}, i.e., not used, for some computation tasks \cite{frozen}. Error mitigation schemes \cite{error1,error2,error3,error4} help to improve computation fidelity \cite{Book,hammer,QNN} but have limited effectiveness due to the limited availability of physical qubits on devices.
As an alternative to physical quantum execution, software simulation offers a noise-free execution environment for quantum circuits. However, the simulation cost increases exponentially with the number of qubits \cite{Large_Simulation} and thus faces intrinsic drawbacks for scalability.

Recently proposed schemes, i.e., {\em wire cutting}, {\em qubit reuse}, and {\em gate cutting}, help to mitigate the challenge. The wire-cutting schemes \cite{cutqc,Peng,clifford_cutting} partition a large circuit into several smaller subcircuits, run subcircuits on quantum devices, and then reconstruct the output of the original circuit through classical post-processing. That is, they adopt a hybrid approach that combines physical quantum execution and classical software post-processing. Since the classical post-processing overhead increases exponentially with the number of wire cuts, minimizing the cut number is the major design goal. The qubit-reuse technique \cite{qubit_reuse,qubitreuse2,qubitreuse3} exploits the hardware support for Mid-Circuit Measurement and Reset (MR)~\cite{mid-circuit} such that the physical qubits that have finished all their operations can be redeployed as other logical qubits during circuit execution. 
Another circuit-cutting approach, gate cutting \cite{Mitarai_2021}, was recently proposed for circuits that compute expectation values, e.g., Hamiltonian simulation algorithms, where minimization of expectation value is the main design goal. Gate cutting cuts a two-qubit gate into a linear sum of single-qubit gates and exploits classical post-processing to reconstruct the original result. Gate cutting has not been well-studied at the circuit level, i.e., deciding the best cut locations for a given large circuit.

Unfortunately, we observe that these schemes are currently applied individually and tend to produce sub-optimal results. 
Qubit reuse can reuse physical qubits only after their initially assigned operations finish. 
Its effectiveness diminishes as the circuits grow larger --- only a few qubits can start their operations after some other qubits have finished. 
Wire cutting introduces one extra qubit ({\em initialization qubit} in~\cite{cutqc}) after each cut, which may artificially increase the total number of physical qubits required for partitioning circuits. 
Gate cutting has not been well-studied at the circuit level. In addition, gate cutting, since its post-processing cannot reconstruct the distribution result, can only be applied to the quantum circuits that compute expectation values. 

In this work, we present \OurScheme{}, a framework for evaluating large quantum circuits on small quantum devices through integrated Qubit-Reuse and circuit-cutting. \OurScheme{} is an end-to-end approach that models a large quantum circuit using ILP (integer linear programming), finds a good cutting solution using an ILP solver, maps the decision to subcircuits,
runs the subcircuits on quantum devices, and reconstructs the original result through classical post-processing.

Compared with prior schemes, our key observation is that wire cuts in the circuit enlarge qubit-reuse opportunities, which in turn help to eliminate unnecessary cuts in the circuit. By integrating qubit reuse and circuit-cutting, \OurScheme{} strives to find better cutting solutions with fewer numbers of cuts, reduced post-processing overhead, and improved per-circuit computation fidelity. 
When only the quantum circuit's expectation value is required, gate cutting can be applied to enlarge the cutting possibilities and thus further decrease the number of required cuts.

We further study in detail, in Section 6.6, the post-processing overhead with regards to i) the number of cuts, ii) the size of the quantum circuit, and iii) the reconstruction strategy for original output construction. We show that the number of cuts required to partition a circuit is the dominant factor contributing to the overhead. We further study the impact of i) circuit and device size, and ii) density of two-qubit gates when partitioning a quantum circuit on the number of cuts required. Our data shows the scalability of our circuit-cutting framework in the NISQ era.

We summarize our contributions as follows.
\begin{itemize}[leftmargin=*]
\itemsep 2pt
    \item We propose \OurScheme{}, a framework for evaluating large circuits on small quantum computers. To the best of our knowledge, \OurScheme{} is the first framework that (i) integrates wire and gate cuttings; and (ii) exploits qubit reuse to take advantage of the opportunities from circuit-cutting.
   
    \item We formulate the problem as a searching framework using ILP, which enables the search for solutions under different optimization goals. The ILP formulation helps to achieve efficiency in an enlarged search space, and better scalability over the state-of-the-art~\cite{cutqc}.
    
    \item We evaluate \OurScheme{} using different benchmarks. Our results show that, on average, we reduce the number of cuts by 29\% when considering wire cutting only and gain additional reduction when considering wire and gate cutting.  We verify our approach using real device execution and post-processing.

    \item We provide a detailed analysis of the post-processing overhead of our framework. We highlight the key factors contributing to the complexity of circuit cutting with respect to the circuit and device size, reconstruction strategy, and cutting strategy.
    
\end{itemize}

\section{Background} \label{background}

\subsection{Quantum Circuits and their Outputs}

A quantum program is represented as a quantum circuit consisting of qubits and quantum gates. Current quantum hardware supports single-qubit and two-qubit gates, which are also the gates considered in this paper.
A quantum circuit, represented as a unitary matrix $\mathcal{U}$, takes an initial qubit state |$\psi\rangle$, usually |0$\rangle$$^{\otimes n}$, and evolves it to an output state |$\phi\rangle$. 
\begin{equation}\label{eq: quantum circuit}
    \mathcal{U}|\psi\rangle = |\phi\rangle
\end{equation}

Many quantum algorithms, e.g., Grover's algorithm for quantum search \cite{grover1996fast}, compute probability vector |$\phi\rangle$, which indicates the probability distribution of measuring each of the possible 2$^n$ states.
Alternatively, other algorithms, such as Variational Quantum Algorithm (VQA) \cite{error1,VQA2}, compute the expectation value of a Hamiltonian in the computational measurement basis M.
\begin{equation}\label{eq: expectation}
    \mathbf{E} = \langle\phi|M|\phi\rangle
\end{equation}



\subsection{Quantum Circuit Simulation}

Quantum circuits can be simulated on classical computers using state-vector simulation. The classical simulation provides an ideal noise-free run of quantum circuits and accurately reproduces the output. However, there is an exponential cost of simulation, which restricts the simulation of larger quantum circuits. Wu {\em et al.} showed that simulating the 61-qubit Grover search algorithm needed Argonne's Theta supercomputer with 4,096 nodes and 768TB memory~\cite{Large_Simulation}.

An alternative approach to simulation is to run the circuits on real quantum computers, using the shots-based model. 
That is, the quantum circuit is executed thousands of times, with each execution referred to as a shot, on the quantum hardware and the measurement of each qubit from each shot is summarized as the output probability vector of the circuit. The drawbacks of this approach are: (i) many shots are required, even in an ideal noise-free setting, to reproduce the output probability vector of the original circuit accurately; (ii) given that today's quantum computers are noisy, the computation fidelity is often low for large circuit execution; (iii) the available quantum computers have limited numbers of qubits. The largest quantum computer from IBM has 433 physical qubits \cite{Osprey}.

\subsection{Circuit-Cutting}
Circuit-cutting is a technique for obtaining the result of a large quantum circuit on small quantum devices. After cutting the large circuit into two or more smaller subcircuits using either {\em wire cutting} or {\em gate cutting}, we can execute subcircuits on small quantum devices, and generate the result of the original circuit from the classical post-processing of the results of subcircuits \cite{Peng}. 

\begin{figure*}[htbp]
\centering
    \subfigure[Wire Cutting]{
        \includegraphics[width=2.24in]{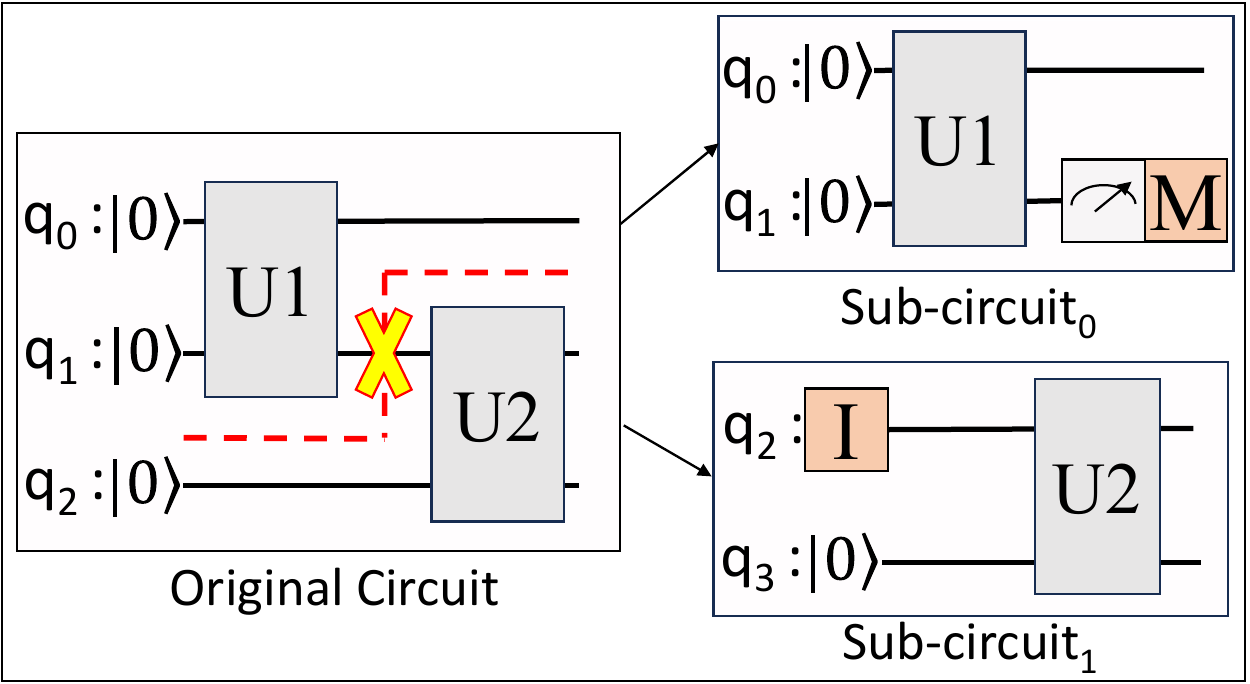}}
    \subfigure[Gate Cutting]{
        \includegraphics[width=2.24in]{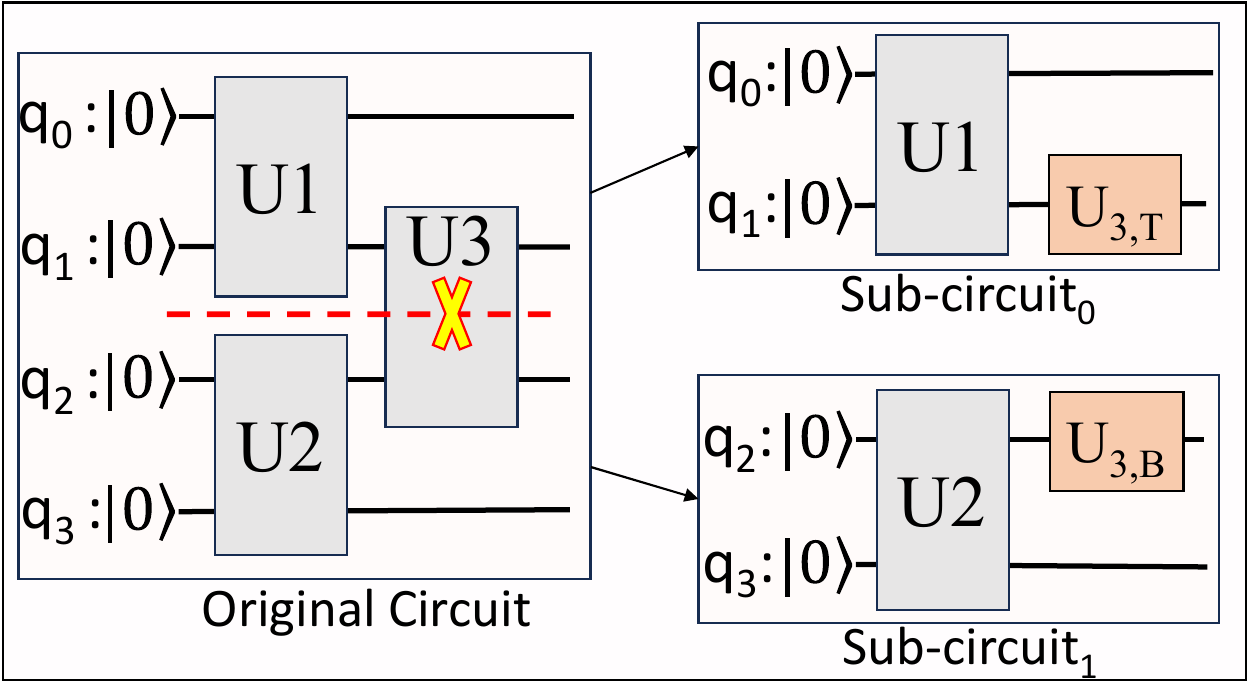}}
    \subfigure[Qubit Reuse]{
        \includegraphics[width=2.24in]{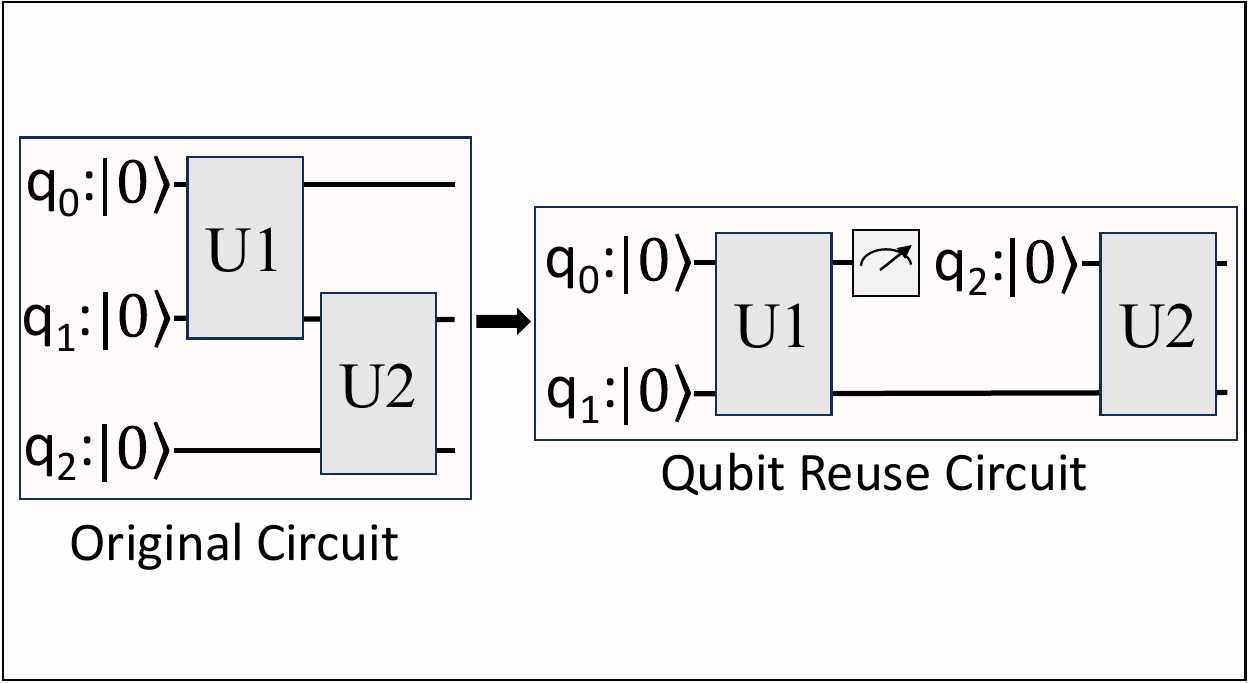}}
    \vspace{-0.15in}
   \caption{Circuit-cutting and qubit reuse. (a) An example of wire cutting is where qubit q$_1$ has been cut. It leaves measurement (M) and Initialization (I) operations in two subcircuits, respectively. (b) An example of Gate cut is where gate U3, acting on qubits q$_1$ and q$_2$ has been cut. It leaves two single-qubit gates in two subcircuits, respectively. (c) An example of qubit reuse. Once the U1 gate has finished executing, qubit q$_0$ can be measured and reused for logical qubit q$_2$.}
   \vspace{-0.12in}
    \label{fig:methods}
\end{figure*}

\subsubsection{Wire Cutting (W-Cut) } \label{wire_cutting_background}
Wire cutting (W-Cut) \cite{Peng,cutqc} cuts the wire that connects two quantum gates, as shown in Figure \ref{fig:methods}(a). Here, U1 and U2 are two generic two-qubit gates. W-Cut cuts the original circuit into two independent subcircuits: {\em subcircuit$_0$} and {\em subcircuit$_1$}, as shown in the figure. To obtain the output state $|\rho\rangle$ of the original circuit, CutQC \cite{cutqc} runs {\em subcircuit$_0$} with measurements in four bases, and {\em subcircuit$_1$} with four initializations; and then reconstructs the output of the original circuit using Equation~(\ref{equation:cutqc}). 
\begin{equation} \label{equation:cutqc}
    \rho = \frac{A_1 + A_2  + A_3 + A_4}{2}
\end{equation}
where
\begin{align}
    A_1 &= [Tr(\rho I) + Tr(\rho Z)] |0\rangle\langle0| \nonumber\\
    A_2 &= [Tr(\rho I) - Tr(\rho Z)] |1\rangle\langle1| \nonumber\\
    A_3 &= Tr(\rho X)[2|+\rangle\langle+| - |0\rangle\langle0| - |1\rangle\langle1|] \nonumber\\
    A_4 &= Tr(\rho Y)[2|i\rangle\langle i| - |0\rangle\langle0| - |1\rangle\langle1|] \nonumber
\end{align}

\noindent
Here, Tr() is the trace operator indicating running {\em subcircuit$_0$} physically on quantum devices and measuring the output in one of the Pauli basis bases (i.e., M $\in$\{I, X, Y, Z\}).
Measuring a qubit in either the I or Z basis gives the same circuit.
$|x\rangle\langle x|$ is the density matrix indicating initializing {\em subcircuit$_1$} in one of the eigen states (i.e., I $\in$\{$|0\rangle$, $|1\rangle$, $|+\rangle$, $|i\rangle$\}). 
From Equation \ref{equation:cutqc}, W-Cut needs four pairs of Kronecker products between the subcircuit results to reconstruct the result of the original circuit.
If it takes $k$ ($k$>0) cuts to partition a large circuit into multiple independent 
subcircuits, the classical post-processing overhead of result reconstruction is O(4$^k$). 

Applying W-Cut at the circuit level is an optimization problem that finds the wires to be cut
in a given large circuit, such that the cutting has the smallest $k$ and ensures the execution of each subcircuit on small quantum devices.
CutQC formulates the problem as an MIP (mixed integer programming) model and exploits an MIP solver to search for the best solution.

\begin{figure*}[tbp]
\begin{center}
\subfigure[The Original Circuit]{\begin{minipage}[b]{0.495\linewidth}
      \includegraphics[width=3.2in]{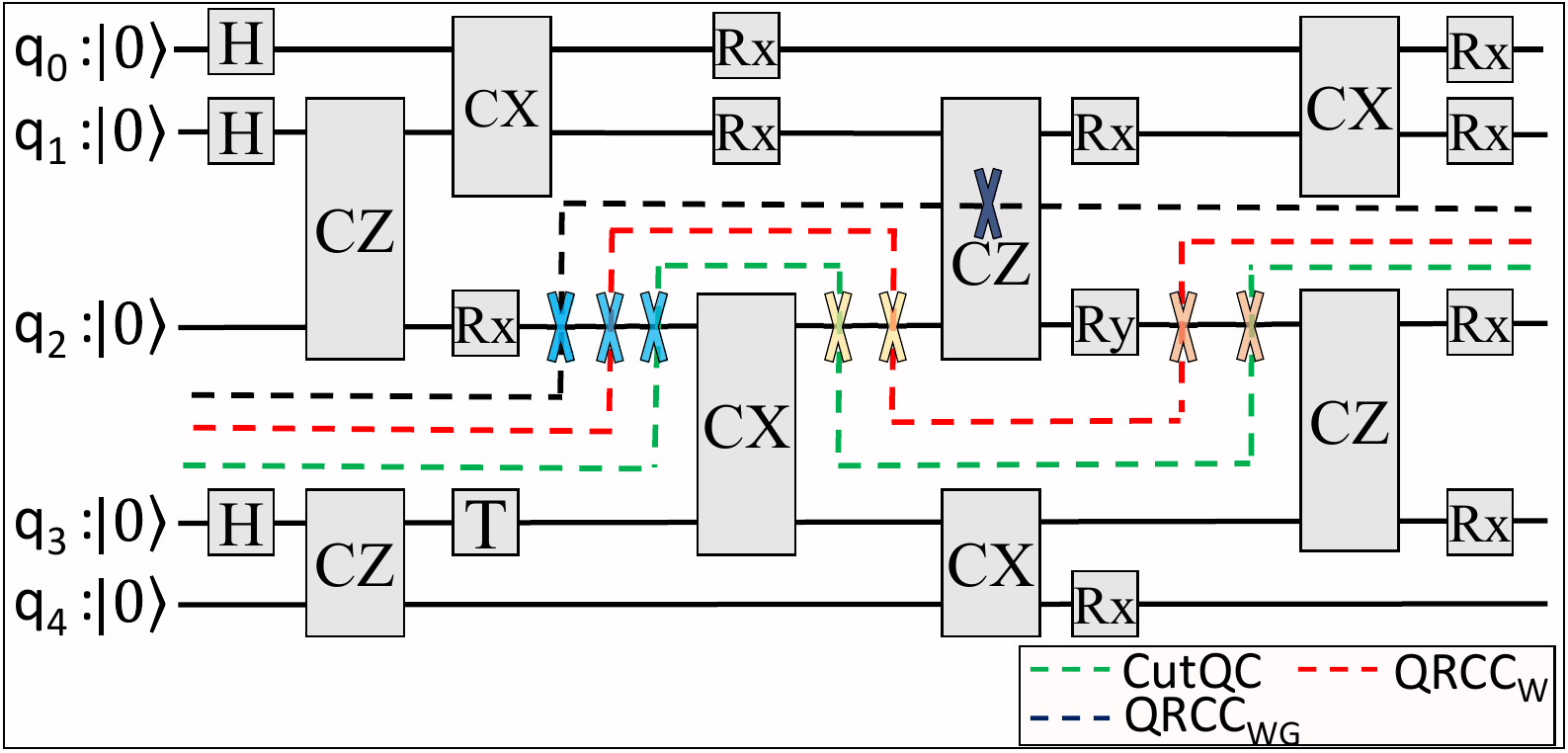}
    \end{minipage}}
\hfill
\subfigure[CutQC generates two 4-qubit subcircuits]{\begin{minipage}[b]{0.495\linewidth}
      \includegraphics[width=3.2in]{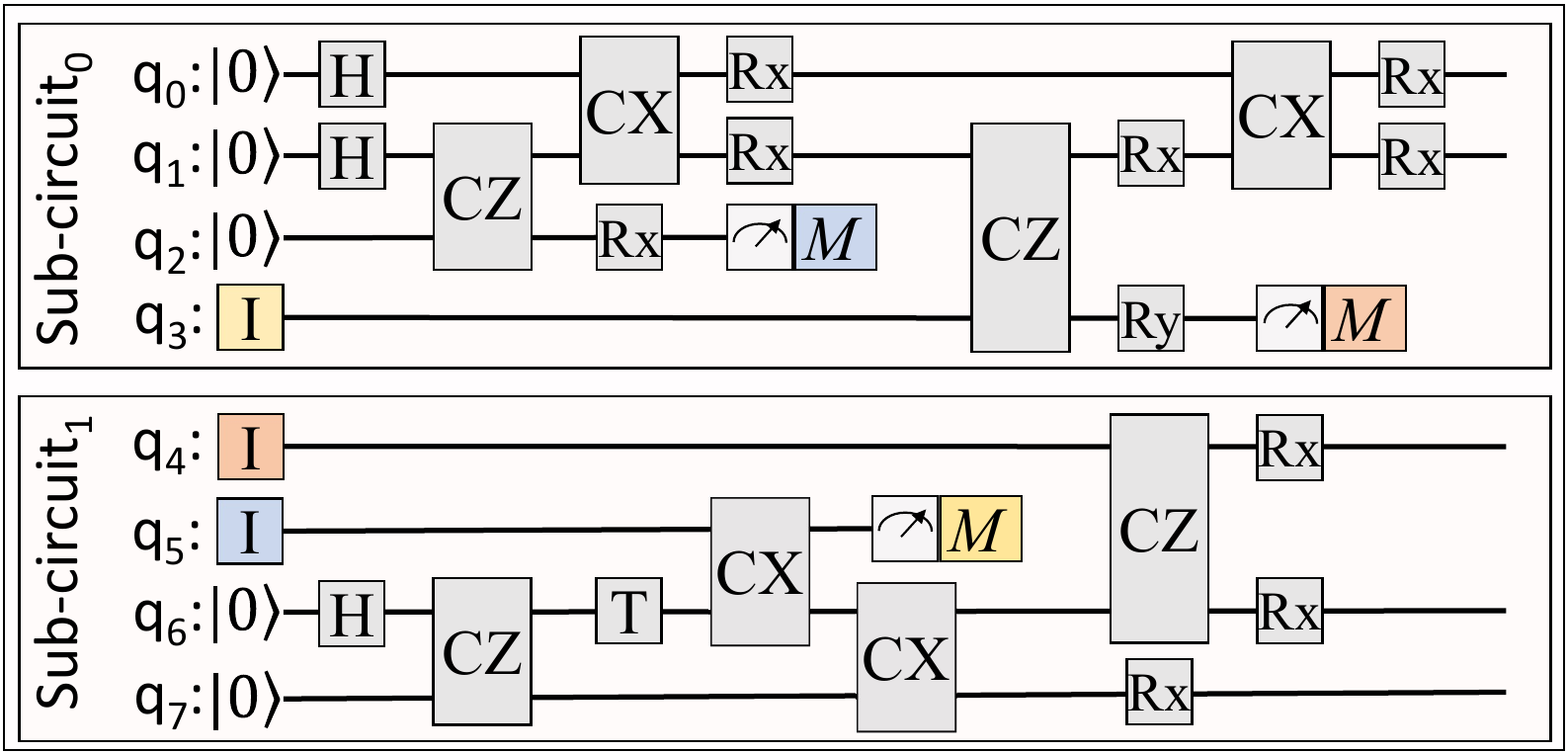}
    \end{minipage}}

\subfigure[The solution when integrating W-Cut and Qubit reuse]{\begin{minipage}[b]{0.495\linewidth}
      \includegraphics[width=3.2in]{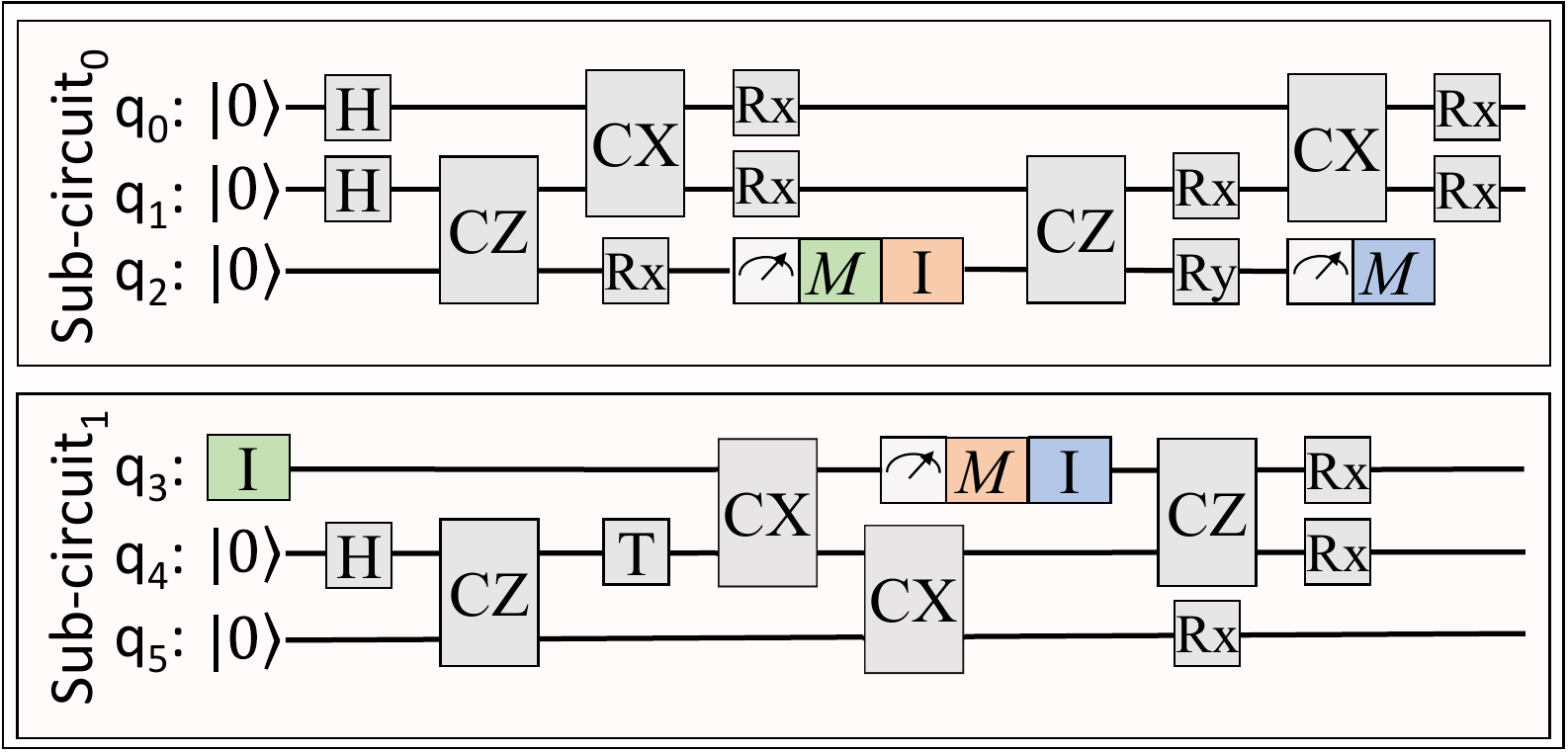}
    \end{minipage}}
\hfill
\subfigure[The solution when integrating W-Cut, G-Cut, and Qubit reuse]{\begin{minipage}[b]{0.495\linewidth}
      \includegraphics[width=3.2in]{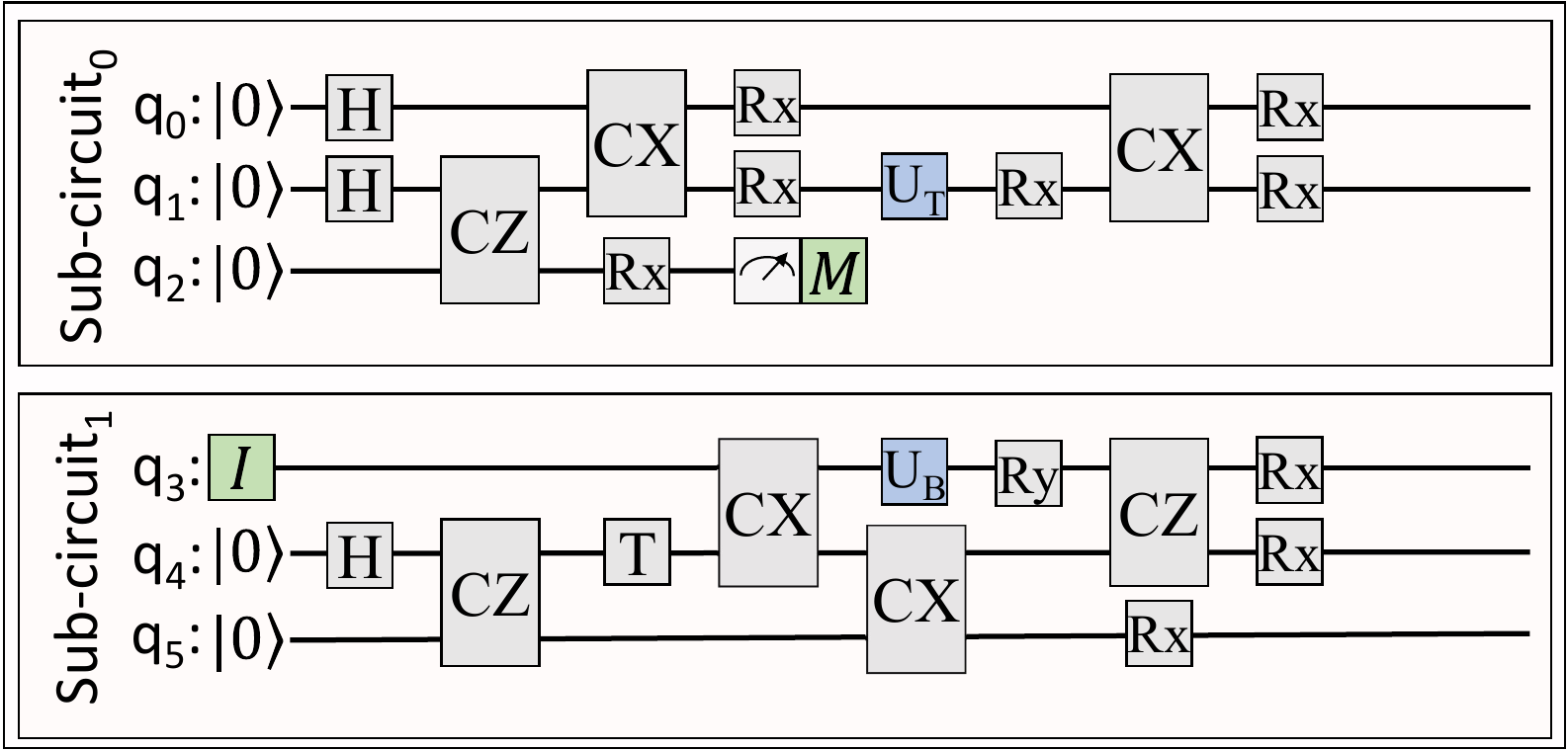}
    \end{minipage}}
\end{center}
\vspace{-0.15in}
 \caption{The integration of W-Cut, G-Cut, and qubit reuse helps to find better cutting solutions. (a) Original Circuit, showing three different cutting solutions. (b) The solution generated by CutQC. (c) The solution when integrating W-Cut and qubit reuse. (d) The solution when integrating W-Cut, G-Cut, and qubit reuse. ({\bf M}/{\bf I} indicate measurement and initialization, respectively, due to W-Cut; and {\bf U$_T$}/{\bf U$_B$} indicate the top/bottom single-qubit gates after applying G-Cut to a two-qubit gate {\bf U}). }
\vspace{-0.05in}
\label{fig: Motivational example}
\end{figure*}

\subsubsection{Gate Cutting (G-Cut) }
\label{gate_cutting_background}
Gate cutting (G-Cut) cuts a two-qubit quantum gate, e.g., $U3$ in Figure \ref{fig:methods}(b), into a linear sum of single-qubit gates $U_{3.T}$ and $U_{3.B}$. According to the theory of gate cutting \cite{Mitarai_2021}, if G-Cut cuts a two-qubit gate of the form $e^{i\theta A_1\otimes A_2}$ (e.g., CNOT, CZ, and ZZ gates) where A$_1^2$ = A$_2^2$ = I, the expectation value $\mathbf{E}$ of the original circuit can be reproduced based on the output state |$\phi_{i}\rangle$ of subcircuits, using Equation (\ref{equation: gate_equation}). 
G-Cut differs from W-Cut in that G-Cut cannot reproduce the original circuit state vector, but rather only the expectation value. 
\vspace{-0.05in}
\begin{equation} \label{equation: gate_equation}
   \mathbf{E}[\phi] = \sum^{6}_{i=1} c_iE[\phi_i]
\end{equation}
\vspace{-0.05in}
where,
\begin{align*}
    \phi_{1} &= \mathbf{S}(I \otimes I) &&c_1 = cos^2(\theta) \\ 
    \phi_{2} &= \mathbf{S}(A_1 \otimes A_2) &&c_2 = sin^2(\theta) \\
    \phi_{3} &= \beta\mathbf{M}_{A_1,\beta}\otimes\mathbf{S}(e^{i\pi A_2/4}) &&c_3 = cos(\theta)sin(\theta)\\
    \phi_{4} &= \beta\mathbf{M}_{A_1,\beta}\otimes\mathbf{S}(e^{-i\pi A_2/4}) && c_4 = -cos(\theta)sin(\theta) \\
    \phi_{5} &= \mathbf{S}(e^{i\pi A_1/4} )\otimes\beta\mathbf{M}_{A_2,\beta} &&c_5 = cos(\theta)sin(\theta) \\
    \phi_{6} &= \mathbf{S}(e^{-i\pi A_1/4} )\otimes\beta\mathbf{M}_{A_2,\beta} &&c_6 = -cos(\theta)sin(\theta) 
\end{align*}

G-Cut produces six subcircuit instances, i.e., $\phi_1$  to $\phi_6$.
Each $\phi_{i}$ is an independent instance, from which, during its execution, we remove the two-qubit gate that has been cut from the original circuit and replace it with single-qubit gates of the respective instance. The M$_{A_i}$ term is single qubit measurement operations, with $\beta$ representing the outcome of the measurement, $\beta$ $\in$ \{1,-1\}. More details can be found in \cite{Mitarai_2021}.

G-Cut has not been well-studied at the circuit level. Given a large circuit, it remains an open problem to determine the subset of two-qubit gates to be cut to achieve our design goal, in particular, together with W-Cut and qubit reuse.

\subsection{Measure and Reset Functionality}

IBM recently introduced mid-circuit measurement operation and mid-circuit reset operation \cite{mid-circuit} to support dynamic circuits for quantum error correction\cite{qec1,qec2,qec3} and runtime program verification\cite{verification,verification2}. As shown in Figure \ref{fig:methods}(c), once qubit q$_0$ finishes its operation with gate $U1$, we measure this physical qubit and re-initialize another logical qubit q$_2$ in the
|0$\rangle$ state, and assign qubit q$_2$ to the same physical qubit on the quantum device.
This is referred to as {\em qubit-reuse} in \cite{qubit_reuse}. In the figure, qubit-reuse enables the execution of the original three-qubit circuit on a two-qubit quantum device. 

CaQR \cite{qubit_reuse} proposes a compiler-assisted tool that automatically identifies qubit-reuse opportunities in a given circuit, reduces the total number of required physical qubits, and achieves better performance and computation fidelity.
The effectiveness of qubit reuse diminishes as the circuit becomes bigger --- only a few qubits can delay their operations enough to start after some other qubits have finished.

\section{Motivation}

In this section, we use an example (in Figure \ref{fig: Motivational example}) to illustrate the effectiveness when integrating qubit reuse, W-Cut, and G-Cut. Our problem is to run a 5-qubit circuit on small quantum devices, e.g., 4-qubit or 3-qubit quantum devices.

When adopting CutQC \cite{cutqc}, the original circuit is split into two subcircuits using three cuts, as shown in Figure~\ref{fig: Motivational example}(b). 
Each subcircuit has four qubits. Three extra qubits (i.e., the initialization qubits) are introduced, e.g., the first wire cut on $q2$ generates an extra qubit $q5$, which is now the second qubit in $subcircuit_1$. It leaves a measurement in $subcircuit_0$.
We use three color pairs to indicate the introduced measurement operation and its matching initialization bit.

For this circuit, CutQC can't find a solution that splits the original circuit into two 3-qubit subcircuits. It is also impossible to apply qubit reuse \cite{qubit_reuse} directly on the original circuit to reduce the number of required qubits.

\subsection{W-Cut and qubit-reuse}
Figure \ref{fig: Motivational example}(c) shows the cutting solution when we integrate W-Cut and qubit reuse. The integrated scheme, even though choosing the same cutting positions as those in CutQC, generates two 3-qubit subcircuits. 

The improvement comes from the reuse opportunities exposed from wire cutting. For example, for $subcircuit_0$, qubit $q2$ becomes idle after the first cut. It can be reused by the initialization qubit. By exploiting the qubit reuse opportunities, each subcircuit requires one fewer qubit so that both can run on three-qubit quantum devices.

W-cut partitions the operations on the cut qubit, such that it introduces new qubit reuse opportunities into the circuit, which previously did not exist.


\subsection{ W-Cut and G-Cut}
Figure \ref{fig: Motivational example}(d) shows the cutting result when we integrate W-Cut and G-Cut. The integrated scheme can cut the original circuit into two subcircuits in two cuts --- one wire cut and one gate cut. The two-qubit gate $CZ$ is cut into two single-qubit gate instances $U_T$ and $U_B$ in different subcircuits. 

In this example, G-Cut is enabled only if the circuit computes the expectation value. 
The classical post-processing overhead also impacts the solution selection, in particular, the cost from G-Cut is slightly higher than that from W-Cut, i.e., $6^k$ vs $4^k$, where $k$ is the number of cuts. Therefore, we need to consider this difference when choosing a cutting solution.
Given a solution S($k1$, $k2)$ where $k1$ and $k2$ are the numbers of gate cuts and wire cuts, respectively, its classical post-processing overhead is O($4^{k1}6^{k2}$). It is better to choose S(1,1) over S(2,1) for the example in the figure. In another situation, it would be worse to choose S(0,4) over S(5,0).

\section{The \OurScheme{} Framework}

In this section, we elaborate \OurScheme{}, an end-to-end framework, for running large quantum circuits on small quantum devices. Given a large circuit, \OurScheme{} converts it to a QR (qubit-reuse)-aware DAG, formulates and solves an ILP model, maps the cutting solutions to subcircuits, runs the subcircuits, and reconstructs the original result.


\subsection{The QR-aware DAG Representation} \label{preprocessing}
Given a $N$-qubit input quantum circuit that is to be cut for an $D$-bit quantum device ($N$>$D$>0), we first convert the circuit to a QR-aware DAG by adding dummy {\em Identity} gates such that, each qubit goes through the same number of quantum operations. 
After adding the identity gates, all qubits are {\em aligned} so that we define a {\bf quantum layer $m$} as the set of $m$-th gate for each qubit.

\begin{figure}[htbp]
    \centering
    \includegraphics[width=2.8in]{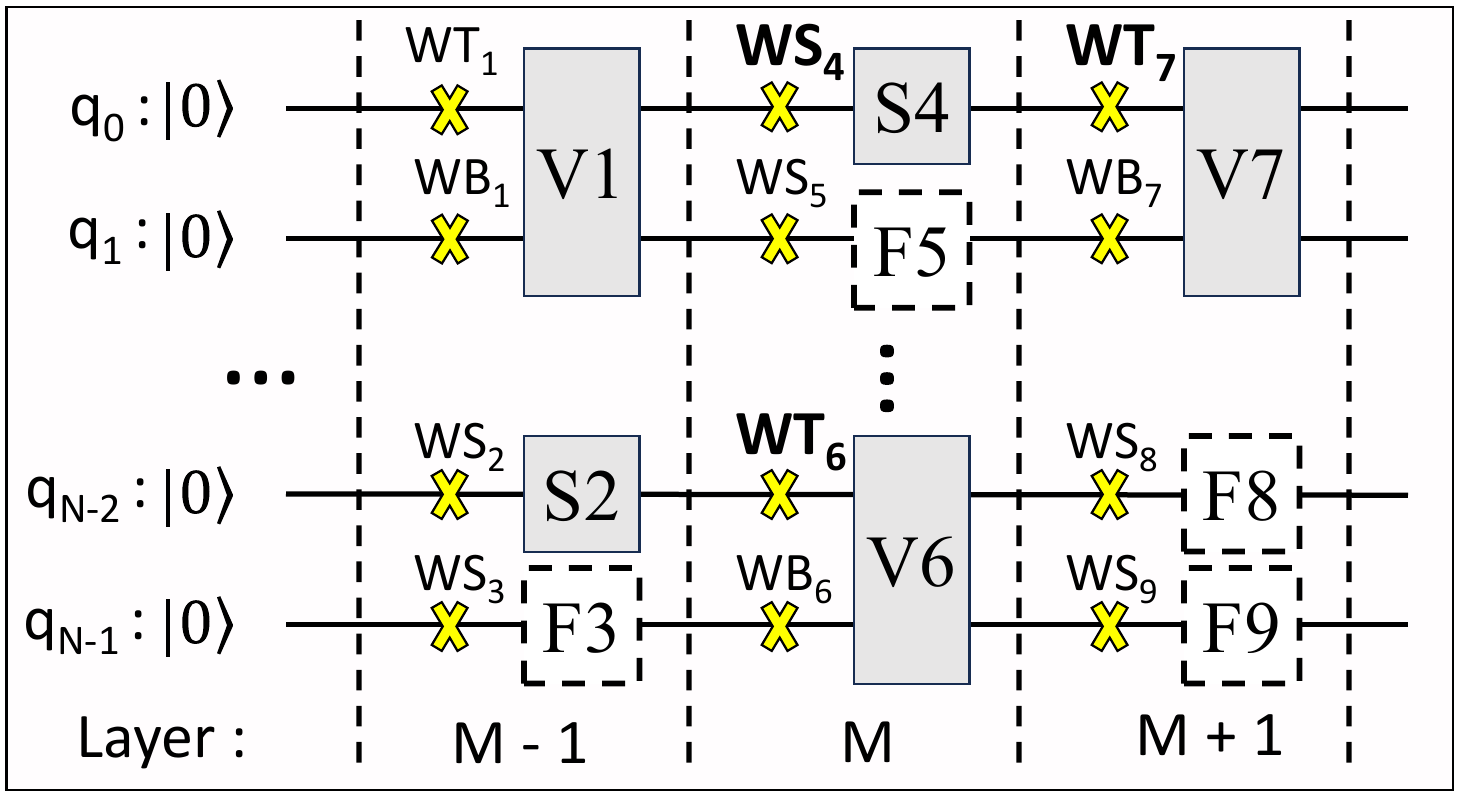}
    \caption{A QR-aware DAG representation of the quantum circuit. (Dashed boxes indicate identity gates. Each wire segment may potentially be W-Cut, and each two-qubit gate may be potentially G-Cut.)}
    \label{fig:DAG Representation}
\end{figure}

In Figure \ref{fig:DAG Representation}, V1, V6, and V7 are two-qubits gates, S2 and S4 are single-qubit gates, and F3, F5, F8, and F9 are inserted {\em Identity} gates. Gates S4, F5, and V6 belong to layer $M$. We place a yellow $\times$ on each wire before a gate to indicate a potential cutting location. 

Compared to the DAG for traditional wire cutting \cite{cutqc} that lists two-qubit gates only, our QR-aware DAG explicitly lists all single-qubit gates and differentiates the cuts on the wires connecting different single qubits.
For example, for the two cuts: (i) $WS_4$ is the one on the wire connecting the S4 gate and (ii) $WT_7$ is the one on the top wire connecting the V7 gate, the traditional DAG treats $WS_4$ and $WT_7$ as the same cut, as cutting at either location does not affect the number of required qubits in each subcircuit. However, when we consider qubit reuse, if there is another cut $WT_6$ (the top wire connecting V6), we may prefer to choose $WS_4$ if gate V1 and V6 are in the same subcircuit.
This is because qubit $q_0$ can be reused for qubit $q_{N-2}$ while cutting at $WS_4$ disables this reuse.
For discussion purposes, we assume the measurement and initialization operations take no depth and qubit-reuse does not increase circuit depth. Section 4.2.6 discusses how to handle the depths of these operations.

\subsection{The ILP Model}
\subsubsection{The Meta Parameters} 
We formulate the problem as an ILP model. We first list the meta parameters, i.e., the constants that we define for the problem and/or we collect from preprocessing the input circuit. 

\begin{itemize}[leftmargin=*]
\item $N$ and $D$: They are the number of qubits in the input quantum circuit and the number of available physical qubits of the quantum device, respectively. We have $N$>$D$>0.
\item $G_{max}$ and $W_{max}$: The maximum number of allowed gate cuts and wire cuts, respectively.
\item $C_{min}$ and $C_{max}$: The minimum and maximum numbers of subcircuits to be cut, respectively.
 Note, that our ILP solver often reports a cutting solution that has fewer than $C_{max}$ subcircuits. This is because our model focuses on reducing the classical post-processing overhead, which does not relate to the number of subcircuits.
 As we elaborate next, the cost relates to a combination of wire cuts and gate cuts, as the two types of cuts have slightly different classical post-processing overhead.

 If $C_{min}$=$C_{max}$, the solution that we find has the specified number of subcircuits.
\item $\delta$: The relative weight for adjusting the optimization goal between classical post-processing overhead and computation fidelity. We will elaborate in Section 4.2.5.
\item $G$-$Cut$-$enabled$: This is a binary parameter indicating if gate cutting should be enabled. As we discussed, after G-Cut, we can only reconstruct the expected value of the circuit. If the original circuit is to compute the probability vector, we disable the gate cutting in the model.
\end{itemize}

\subsubsection{ILP Variables}
When preprocessing the original circuit, we differentiate three types of gates, i.e., two-qubit gates, single-qubit gates in the original circuit, and identity gates that we inserted. We number all gates and define a binary variable for each of the gates as follows.
\begin{eqnarray}
     V_{x,c} &=&
        \begin{cases}
            1 & \text{if two-qubit gate $x$ is in subcircuit $c$}\\
            0 & \text{Otherwise}
        \end{cases} \nonumber \\ 
     S_{x,c} &=&
        \begin{cases}
            1 & \text{if single-qubit gate $x$ is in subcircuit $c$}\\
            0 & \text{Otherwise}
        \end{cases}  \\  
     F_{x,c} &=&
        \begin{cases}
            1 & \text{if identity gate $x$ is in subcircuit $c$}\\
            0 & \text{Otherwise}
        \end{cases} \nonumber
\end{eqnarray}

For single-qubit and identity gates, we can only perform W-Cut. We set the cutting point on the wire before each gate. We do not Cut any gate on the first layer.
\begin{equation}
     WS_{x} = 
        \begin{cases}
            1 & \text{if single-qubit/identity gate $x$ is W-Cut,}\\
            0 & \text{Otherwise}
        \end{cases}
\end{equation}

For two-qubit gates, we can perform both W-Cut and G-Cut. For W-Cut, we can cut either of its input wires. We define the following variables. 

\begin{eqnarray}
     U_{x} &=& 
        \begin{cases}
            1 & \text{if two-qubit gate x is neither W-Cut}\\
              & \text{nor G-Cut}\\
            0 & \text{Otherwise}
        \end{cases}  \nonumber\\
     WT_{x} &=&  
        \begin{cases}
            1 & \text{if top wire to two-qubit gate $x$}\\
              & \text{is W-cut}\\
            0 & \text{Otherwise}
       \end{cases} \\
    WB_{x} &=&  
        \begin{cases}
            1 & \text{if bottom wire to two-qubit gate $x$}\\
              & \text{is W-cut}\\
            0 & \text{Otherwise}
        \end{cases}  \nonumber \\
    G_{x} &=& 
        \begin{cases}
            1 & \text{if two-qubit gate x is G-Cut}\\
            0 & \text{Otherwise}
        \end{cases}  \nonumber
\end{eqnarray}

When G-Cutting a two-qubit gate $x$, we get two single-qubit gates, referred to as $x.top$ and $x.bottom$. These two gates appear only if $G_x$=1.
Similar to those in definition (6), we define variables to determine if they are in some subcircuits.
\begin{eqnarray} 
    GT_{x,c} &=&
        \begin{cases}
            1 & \text{if for two-qubit gate $x$, we have } \\
              & \text{$G_x$=1 and $x.top$ is in subcircuit c}\\
            0 & \text{Otherwise}
        \end{cases} \\
    GB_{x,c} &=&
        \begin{cases}
            1 & \text{if for two-qubit gate $x$, we have } \\
              & \text{$G_x$=1 and $x.bottom$ is in subcircuit c}\\
            0 & \text{Otherwise}
        \end{cases}  \nonumber       
\end{eqnarray}

\subsubsection{The General ILP Constraints} 
We next list the general constraints in our model. These constraints are the same regardless of the input circuit and user parameters.   

Whether a single-qubit or identity gate $x$ is cut is determined by its $WS_x$ variable. However, a two-qubit gate may be W-Cut, G-Cut, or not cut. That is, we cannot W-Cut and G-Cut the gate at the same time. Therefore, we have the following constraints for each two-qubit gate $x$.
\begin{eqnarray}
U_x + WT_x + WB_x + G_x &\geq& 1    \nonumber \\
U_x + WT_x &\leq& 1     \nonumber \\
U_x + WB_x &\leq& 1     \\ 
U_x + G_x &\leq& 1     \nonumber  
\end{eqnarray}
Each gate $x$ must belong to one and only one subcircuit, unless it is a two-qubit gate and has been G-Cut. 
If a two-qubit gate $x$ is G-Cut, its two-qubit gate form $conceptually$ disappears such that the two single-qubit gates, i.e., $x.top$ and $x.bottom$, emerge in the circuit. In this case, the newly generated single-qubit gates, i.e., $x.top$ and $x.bottom$, must belong to one and only one subcircuit. These two single-qubit gates cannot belong to the same subcircuit.

We use this technique to linearize the gate cut constraints. This technique is also used in Section 4.2.6.

\begin{eqnarray}
\text{for single-qubit gate $x$, }\sum_{c\in C}S_{x,c} &=& 1  \nonumber\\
\text{for two-qubit gate $x$, }\sum_{c\in C}V_{x,c} + G_x &=& 1 \\
\sum_{c\in C}GT_{x,c}  &=& G_x \nonumber \\
\sum_{c\in C}GB_{x,c}  &=& G_x \nonumber \\
\text{for $\forall$ subcircuit $c\in C$,~~~~~}   GT_{x,c} + GB_{x,c} &\leq& 1 \nonumber 
\end{eqnarray}

After cutting, each subcircuit should not have more than $D$ qubits, i.e., the device size constraint. Therefore, for all identity gates $x$, single-qubit gates $s$, and two-qubit gates $t$, respectively at each layer $l$, 
\begin{eqnarray}
Q_{c,l} = \sum_{x}F_{x,c} + \sum_{s}S_{s,c} + \sum_{t}(2V_{t,c} + GT_{t,c} + GB_{t,c}) \leq D
\end{eqnarray}
\noindent
where $Q_{c,l}$ is the number of qubits used in subcircuit $c$ at layer $l$. By adopting the layer-based cutting approach, our model allows us to find better qubit reuse opportunities such that a wire cut at an early layer can be reused by a different qubit at a later layer.

We also restrict the number of gate cuts and wire cuts.
\begin{eqnarray}
    \sum_{x} G_x   &\le& G_{max}  \nonumber \\
    \sum_{x} (WS_x + WT_x + WB_x) &\le& W_{max}
\end{eqnarray}

\subsubsection{The Circuit-dependent Constraints} \label{conditional_constraints_sections}
In addition to the general constraints, we have circuit-dependent constraints. These constraints specify the relationship between two neighboring gates.

If two neighboring gates are two two-qubit gates, we may have two cases: (a) the {\underline{bottom}} output of the upstream gate connects to the {\underline{top}} input of the downstream gate, e.g, the U1-U3 connection in Figure \ref{fig:methods}(b); or (b) the {\underline{top}} output of the upstream gate connects to the {\underline{bottom}} input of the downstream gate, e.g, the U2-U3 connection. We specify their constraints as follows.
\begin{eqnarray}
     2\times WT_{U3} &= \sum_{c\in C}(|V_{U1,c} - V_{U3,c} + GB_{U1,c} - GT_{U3,c}|) \nonumber \\
     2\times WB_{U3} &= \sum_{c\in C}(|V_{U2,c} - V_{U3,c} + GT_{U2,c} - GB_{U3,c}|)  
\end{eqnarray} 

If two neighboring gates are one upstream single-qubit gate and one downstream two-qubit gate, for example, if we 

\noindent
replace U1 and U3 with single-qubit gates, the constraints are 
\begin{eqnarray}
     2\times WT_{U3} &= \sum_{c\in C}(|S_{U1,c} - V_{U3,c} - GT_{U3,c}|) \nonumber \\
     2\times WB_{U3} &= \sum_{c\in C}(|S_{U2,c} - V_{U3,c} - GB_{U3,c}|)  
\end{eqnarray} 

Similar constraints are specified for other circuit connections. They follow the same rules of gate and wire cuts. 

\subsubsection{The Objective Function} 
Our ILP model consists of two optimization goals. 
\begin{itemize}[leftmargin=*]
\item Our main optimization goal is to reduce the number of cuts to minimize the classical post-processing overhead. 

Since the overhead of a cutting solution ($k$,$m$), i.e., with $k$ wire cuts and $m$ gate cuts, is O(4$^k$6$^m$),
a naive integration of this overhead in the objective function would lead to a non-linear component, which can greatly slow down the solver. Instead, we linearize the cost as $\alpha k + \beta m$  such that if 
the exponential cost of ($k_1$,$m_1$) is smaller than that of ($k_2$,$m_2$), our linear cost has the same relative relationship. In this work, we choose $\alpha$=3.25 and $\beta$=4.2 as they satisfy the requirement for the number of cuts smaller than 240 (120 W-cut and 120 G-cut). 

Therefore, the classical post-processing overhead is 
\begin{equation} \label{objective} 
    PPCost = \alpha\times \sum_{x}(WS_x + WT_x + WB_x) + \beta\times \sum_{x} G_x
\end{equation} 

\item The other optimization goal of our model is to improve the computational fidelity.
Studies have shown that the computation error of a quantum circuit depends on the number of operations, in particular, two-qubit quantum operations \cite{gate-error1,gate-error-2,gate-error-3}. This is because the error rate of two-qubit gates is orders of magnitude higher than that of single-qubit gates. To improve the computational fidelity after circuit cutting, we strive to balance the number of two-qubit gates across different subcircuits. 

We define a new variable $TE$ to track the maximal number of two-qubit gates in a subcircuit. Minimizing $TE$ would help to improve the overall computation fidelity. We add one linear constraint for each subcircuit $c$ as follows. This helps to find the subcircuit that has the maximal number of two-qubit gates.
\begin{align}
    TE &\ge \sum_x V_{x,c} 
\end{align}
We further define the two-qubit gate-related error as 
\begin{equation} \label{objective2} 
    CError = f(TE)
\end{equation} 

Next, we illustrate how to choose a linear function so that $PPCost$ and $CError$ have similar value ranges. 
We use an example to explain how to choose the function. We first run the model considering $PPCost$ only such that we may find a cutting solution (4, 6) with the PPCost value being 3.25$\times$4+4.2$\times$6=38, the number of subcircuits being~5, and the current TE being 40. Assuming we can achieve a perfect balancing of the subcircuits with a maximal increase of 4 additional cuts and get a solution (6,7) with $PPcost$ being 53. The PPCost range is [38,53]. The TE range is now [20,40]. We choose a linear function $f$(TE)= TE$\times$0.75+23.
Note that a further refined linear function can be derived for a given circuit.
\end{itemize}

\noindent
Oftentimes, balancing the number of two-qubit gates across subcircuits may result in a cutting solution with more cuts and thus higher classical post-processing overhead, or even no solution. Therefore, we introduce another meta parameter $\delta$ to adjust the optimization goal between $PPCost$ and $CError$. The $\delta$ value can be integrated into deciding the linear function in $CError$.

To summarize, our objective function is as follows.
\begin{equation} \label{objective3} 
    \textbf{Min}[ \delta \times PPCost + (1-\delta)\times CError]
\end{equation}


\subsubsection{Discussion} 
We make two simplifications for clarity purposes in the preceding discussion of the model.
(1) We assume adding $Identity$ gates to ensure all layers have $N$ gates. This may introduce a large number of identity gates and their corresponding constraints, which slows down the solver. In our implementation, for a long wire that connects two gates far away from each other, we selectively add two or three identity gates at the beginning, middle, and end of the wire.
(2) We assume the measurement and initialization operations take no depth. However, when considering their depths, we introduce a trailing measurement gate after the cut point and a leading initialization gate before the cut point. These two gates emerge in the circuits only if the corresponding wire is cut. We use the same technique as that for gate cutting, i.e., only if a two-qubit gate is cut, its corresponding single-qubit gates emerge in the circuit.

\subsection{Output Reconstruction} \label{reconstruction}

\paragraph{Reconstruction after W-Cut.}
If the original quantum circuit computes the probability distribution vector, we can only adopt wire-cut. The probability vector results from the subcircuit runs can be recombined using Equation (\ref{equation:cutqc}). 
The classical post-processing process follows the techniques as elaborated in CutQC~\cite{cutqc}. 

\paragraph{Reconstruction after W-Cut and G-Cut.}
The original quantum circuit, if computing the expectation value, can be cut by both W-Cut and G-Cut. The reconstruction overhead of expectation values is lower than that of probability vectors, as the expectation value is a floating-point value, while a probability vector consists of multiple floating-point values.

To reconstruct the expectation value of the original circuit, we sort the subcircuits according to the number of their qubits and then start from the smallest subcircuit.
For each subcircuit, we first reconstruct at the wire-cutting positions and then at the gate-cutting positions. 
When handling the W-Cut wires, we reconstruct the expectation values directly, instead of the probability vectors. The Equation (\ref{equation: recompose_A_cutQC}) in Section 2 is applicable for both probability vectors and expectation values \cite{Peng}. For the latter, it can be adapted as follows.
\begin{equation} \label{equation: recompose_A_cutQC}
    \mathbf{E}[\rho] = \frac{A_1 + A_2  + A_3 + A_4}{2}
\end{equation}
where
\begin{align*}
    A_1 &= [\mathbf{E}[Tr(\rho I)] + \mathbf{E}[Tr(\rho Z)]](\mathbf{E}[|0\rangle\langle0|]) \\
    A_2 &= [\mathbf{E}[Tr(\rho I)] - \mathbf{E}[Tr(\rho Z)]](\mathbf{E}[|1\rangle\langle1|]) \\
    A_3 &= \mathbf{E}[Tr(\rho X)](2\mathbf{E}[|+\rangle\langle+|] - \mathbf{E}[|0\rangle\langle0|] - \mathbf{E}[|1\rangle\langle1|]) \\
    A_4 &= \mathbf{E}[Tr(\rho Y)](2\mathbf{E}[|i\rangle\langle i|] - \mathbf{E}[|0\rangle\langle0|] - \mathbf{E}[|1\rangle\langle1|]) 
\end{align*}

After the reconstruction from W-Cut, we adopt Equation~(\ref{equation: gate_equation}) to handle G-Cut for reconstructing the expectation value of the original circuit.

\begin{figure}[tbp]
    \centering
    \includegraphics[width=3.25in]{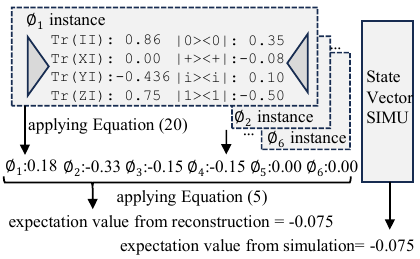}
    \caption{The reconstruction of the expectation value after W-Cut and G-Cut.}
    \label{fig:reconstruction}
\end{figure}

\noindent
\paragraph{An example.}
We next illustrate the reconstruction process for the example shown in Figure \ref{fig: Motivational example}(d), which consists of both W-Cut and G-Cut. For Figure \ref{fig: Motivational example}(c) that contains only W-Cut, the reconstruction process is the same as that in CutQC\cite{cutqc}.

In Figure \ref{fig: Motivational example}(d), the original quantum circuit was cut into two subcircuit, with one wire cut and one gate cut on a CZ gate. The CZ gate has the form
\begin{equation} \label{eq: CZ gate}
    CZ = e^{\frac{i\pi I\otimes Z}{4}}e^{\frac{i\pi Z\otimes I}{4}}e^{\frac{i\pi Z\otimes Z}{4}}
\end{equation}

Each exponential term in this form can be decomposed as shown in Equation (\ref{equation: gate_equation}). We can then combine and simplify all three terms to the following six instances \cite{Mitarai_2021}. These instances are independent of each other.
\begin{align*}
    \phi_{1} &= \mathbf{S}(rz(\frac{-\pi}{2}) \otimes rz(\frac{-\pi}{2})) &&c_1 = \frac{1}{2} \\ 
    \phi_{2} &= \mathbf{S}(rz(\frac{\pi}{2}) \otimes rz(\frac{\pi}{2})) &&c_2 = \frac{1}{2} \\
    \phi_{3} &= \beta\mathbf{M}_{Z,\beta}\otimes\mathbf{S}(e^{i\pi Z/2}) &&c_3 = \frac{-1}{2}\\
    \phi_{4} &= \beta\mathbf{M}_{Z,\beta}\otimes\mathbf{S}(I) && c_4 = \frac{1}{2} \\
    \phi_{5} &= \mathbf{S}(e^{i\pi Z/2} )\otimes\beta\mathbf{M}_{Z,\beta} &&c_5 = \frac{1}{2} \\
    \phi_{6} &= \mathbf{S}(I )\otimes\beta\mathbf{M}_{Z,\beta} &&c_6 = \frac{-1}{2} 
\end{align*}

For example, during $\phi_1$ instance's execution, we replace the two-qubit CZ gate with two single-qubit $rz(\frac{-\pi}{2})$ gates. We then reconstruct the expectation value of this instance, for the wire cut (whose reconstruction follows Equation (20)). Figure \ref{fig:reconstruction} shows the reconstructed expectation value of each $\phi_i$ (1$\le$i$\le$6) using Equation (20) and then the expectation value of the original circuit using Equation (\ref{equation: gate_equation}).
For verification purposes, the expectation value of the original circuit is also computed through state vector simulation, which shows the same result.

\section{Experimental Methodology}
We evaluate the effectiveness of \OurScheme{} using different benchmarks and compare the results with those from CutQC \cite{cutqc}, the state-of-the-art wire-cutting scheme. While both were implemented using the Gurobi optimizer\cite{gurobi}, \OurScheme{} builds an ILP (integer linear programming) model while CutQC builds an MIP (mixed integer programming) model. 
In the experiments, we set the maximal number of cuts to 100. The imbalance threshold is set as 500 gates for CutQC. 
In the experiments, we choose two $\delta$ values for case study purposes and present a comprehensive study of the meta-parameter in Section 6.4.
\begin{itemize}[leftmargin=*]
    \item {\bf \OurScheme{}-C:} we choose $\delta$=1, i.e., we focus on post-processing overhead only. The subcircuits, due to their smaller sizes, generally have better computational fidelity than that of the original circuit. However, some subcircuits may have significantly better computational fidelity than others if they contain fewer two-qubit gates.
    \item {\bf \OurScheme{}-B:} we choose $\delta$=3/4 such that 
    in addition to the main design goal of reducing the post-processing overhead, we also balance the two-qubit gates for computational fidelity improvement.
\end{itemize}
We also run experiments on the IBM Lagos quantum computer through the IBM cloud service to verify our approach.

\begin{table*}[h!]
  \centering
    \footnotesize
    \vspace{-0.1in}
  \caption{Comparing W-Cut results using \OurScheme{} and CutQC. (
  D and N are meta parameters in Section 4.2.1; 
  \#SC: the number of subcircuits after cutting; 
  \#Cuts: the number of wire cuts;
  \#MS: the maximal number of two-qubit gates in the subcircuits)}
  \begin{tabular}{||p{2em}|m{2em}|m{2em}||m{2em}|m{2em}|m{2em}||m{2em}|m{2em}|m{2em}||m{2em}|m{2em}|m{2em}||}
    \hline
    \multicolumn{3}{||c||}{Benchmark} & \multicolumn{3}{c||}{CutQC} & \multicolumn{3}{c||}{\OurScheme{}-C} & \multicolumn{3}{c||}{\OurScheme{}-B}\\ \hline
       & N & D  & \#SC & \#cuts & \#MS & \#SC & \#cuts & \#MS & \#SC & \#cuts & \#MS\\ \hline \hline
       
    \multirow{6}{3em}{QFT} & 15 & 7 & \multicolumn{3}{c||}{No Solution} & 3 & 20 & 69 & 3 & 20 & 68\\
    \cline{2-12}
    & 15 & 9 & 9 & 44 & 27 & 2 & 12 & 81 & 2 & 12 & 75\\  
    \cline{2-12}
    & 30 & 16 & \multicolumn{3}{c||}{No Solution} & 2 & 28 & 330 & 2 & 28 & 318\\  
    \cline{2-12}
    & 30 & 20 & \multicolumn{3}{c||}{No Solution} & 2 & 20 & 380 & 2 & 20 & 335\\  
    \cline{2-12}
    & \multirow{2}{2em}{30} & \multirow{2}{2em}{24} & \multirow{2}{2em}{4} & \multirow{2}{2em}{52} & \multirow{2}{2em}{276} & 2 & 12 & 414 & 2 & 12 & 399 \\ 
    \cline{7-12}
    & & & & & & 4 & 14 & 413 & 4 & 32 & 145 \\ 
    \cline{2-12}
    & \multirow{2}{2em}{30} & \multirow{2}{2em}{27} & \multirow{2}{2em}{3} & \multirow{2}{2em}{32} & \multirow{2}{2em}{351} & 2 & 6 & 429 & 2 & 6  & 426\\ 
    \cline{7-12}
    &    &    &   &    &     & 3 & 7 & 428 & 3 & 30 & 146\\  
    \hline

    
    \multirow{4}{3em}{SPM} & 15 & 7 & 3 & 6 & 8 & 3 & 5 & 9 & 3 & 6 & 8\\
    \cline{2-12}
    & 20 & 7 & 5 & 11 & 8 & 4 & 9 & 13 & 4 & 9 & 9\\
    \cline{2-12}
    & 30 & 16 & 3 & 8 & 22 & 2 & 6 & 25 & 2 & 6 & 25\\
    \cline{2-12}
    & 42 & 16 & 4 & 13 & 21 & 3 & 12 & 26 & 3 & 12 & 24\\
    \hline

    \multirow{4}{3em}{ADD} & 16 & 7 & 4 & 6 & 35 & 3 & 4 & 51 & 3 & 4 & 51\\
    \cline{2-12}
    & 22 & 7 & 5 & 8 & 34 & 4 & 6 & 51 & 4 & 6 & 51\\
    \cline{2-12}
    & 30 & 16 & 2 & 2 & 120 & 2 & 2 & 120 & 2 & 2 & 120\\
    \cline{2-12}
    & 40 & 16 & 3 & 4 & 119 & 3 & 4 & 120 & 3 & 4 & 119\\
    \hline 

    \multirow{4}{3em}{AQFT} & 15 & 7 & 4 & 10 & 18 & 4 & 10 & 18 & 4 & 10 & 16\\
    \cline{2-12}
    & 20 & 7 & 7 & 22 & 20 & 6 & 22 & 31 & 6 & 22 & 19\\
    \cline{2-12}
    & 30 & 16 & 3 & 8 & 65 & 3 & 8 & 65 & 3 & 8 & 65\\
    \cline{2-12}
    & 40 & 16 & 4 & 16 & 74 & 4 & 16 & 75 & 5 & 16 & 71\\
    \hline 
    \end{tabular}
  \label{table:Wire_cut_only_results}

\end{table*}

\subsection{Benchmarks}
We test our scheme using two groups of benchmarks: one computes the probability distribution while the other computes the expectation value. We generate multiple quantum circuits for each benchmark.
We use a three-letter abbreviation to indicate each benchmark, the acronym is in the parameters as we describe each benchmark next.

The following four benchmarks compute the probability distribution and thus can only be cut using W-Cut.
\begin{itemize}[leftmargin=*]
\item\textbf{QFT (QFT)}: Quantum Fourier Transform \cite{qft} is an important building block in many quantum algorithms, including Shor's factoring algorithm. 
\item\textbf{AQFT (AQFT)}: Approximate Quantum Fourier Transform is an approximation \cite{aqft} of the QFT sub-routine, which tends to produce better results on NISQ devices.
\item\textbf{Supremacy (SPM)}: This is a type of random circuit that was used by Google to demonstrate quantum supremacy~\cite{Supremacy}. 
\item\textbf{Adder (ADD)}: This is a linear Ripple Carry Adder~\cite{adder}, which reduces the number of required ancilla qubits  to~1. 
\end{itemize}

To evaluate the effectiveness of G-Cut together with W-Cut, we 
choose the following five variational quantum algorithms that compute expectation values.

\begin{itemize}[leftmargin=*]
    \item\textbf{m-Regular (REG):} The graph in REG is a regular graph in which each node has $m$ edges \cite{n-regular}. By default, $m$=5.
    \item\textbf{Erdos-Renyi (ERD):} The graph in ERD is a random graph in which we exploit a probability $p$ in creating edges across different nodes in the graph \cite{erdos}. By default, $p$=0.1.
    \item\textbf{Barabasi-Albert (BAR):} The graph for this benchmark is also a random graph. Each node in the graph has $m$ edges that connect preferentially to nodes with high degrees \cite{bara}. by default, $m$=3.
    \item\textbf{Hamiltonian Simulation}: For the 2D square lattice Hamiltonian simulation \cite{hamiltonian_sim}, we choose three variations: 2D Traverse Field Isling (IS), XY(XY), and Heisenberg(HS) Hamiltonian. For each variation {\tt M}, 
    we use {\tt M} and {\tt M-n} to indicate the interactions for the nearest neighbor and the next nearest neighbor, respectively.
    \item \textbf{Variational Quantum Eigensolver (VQE)}: We simulate the Hydrogen chain VQE algorithm, using a linear two-local ansatz \cite{VQE}.
    
\end{itemize}

\section{Experimental Results}  

\subsection{Wire Cutting Evaluation}
Table \ref{table:Wire_cut_only_results} compares the W-Cut only results when adopting three different cutting schemes, i.e., CutQC, QRCC-C, and QRCC-B, on benchmarks that compute probability vectors. We report the number of subcircuits (\#SC), the required number of W-Cuts (\#cuts), and the number of two-qubit gates in the largest subcircuit (\#MS). When CutQC cannot find a solution, we report {\em no-solution} in the table. 

Given two cutting solutions {\tt C1} and {\tt C2}, if {\tt C1} cuts the original circuit to fewer subcircuits than {\tt C2} does, {\tt C1}'s subcircuits tend to have more two-qubit gates such that  {\tt C1} tends to have larger \#MS values and thus worse computational fidelity. 
For a fair comparison, if QRCC cuts the original circuit to fewer subcircuits with fewer numbers of cuts than CutQC does, we also report the solutions from QRCC that, with slightly more cuts, cut the original circuits into the same number of subcircuits as CutQC does. For example, for QFT(N=30, D=27), CutQC cuts the original circuits to four subcircuits while, by default, QRCC-B cuts into two subcircuits, resulting in larger \#MS and worse computational fidelity than that of CutQC, i.e., 351 in CutQC vs 426 in QRCC-B. However, if four subcircuits are allowed, QRCC-B achieves better-balanced subcircuits, i.e., 351 in CutQC vs 146 in QRCC-B.


From the table, our scheme significantly reduces the number of cuts ---  on average, \OurScheme{}-C and \OurScheme{}-B achieve 29\% and 24\% reductions over CutQC, respectively.
The test cases from QFT have the most complicated circuits, i.e., more two-qubit gates that exhibit all-to-all qubit connections. 
For these test cases, CutQC may not find a solution if the device size $D$ is small; and \OurScheme{} achieves the largest improvements. For example, \OurScheme{} reduces the \#cuts from 32 to 6 when N=30 and D=27, exhibiting 81\% improvement. As a comparison, for the AQFT benchmark that approximates QFT with all-to-all connections removed, it is easier to find a cutting solution, resulting in negligible improvements over CutQC.


\begin{table*}[h!]
  \centering
    \footnotesize
  \caption{Comparison of W-Cut and W-Cut+G-Cut schemes (\#EffCuts is the effective wire-cuts for comparison) for benchmarks computing expectation values.}
  \begin{tabular}{||p{2em}|m{2em}|m{2em}||m{2em}|m{2em}|m{3em}||m{2em}|m{2em}|m{3em}||m{2em}|m{3.5em}|m{3.5em}|m{3em}|m{3em}||}
    \hline
    \multicolumn{3}{||c||}{Benchmark} & \multicolumn{3}{c||}{CutQC} & \multicolumn{3}{c||}{\OurScheme{}-C (W-Cut Only)} & \multicolumn{5}{c||}{\OurScheme{}-C (W-Cut and G-Cut)}\\
    \hline
     & N & D & \#SC & \#Cuts & \#MS & \#SC & \#Cuts & \#MS & \#SC & \#W-cuts & \#G-cuts & \#EffCuts & \#MS\\  \hline \hline
    \multirow{2}{3em}{REG} & 40 & 27 & 3 & 21 & 49 & 2 & 17 & 51 & 2 & 15 & 1 & 16.29  & 55\\ \cline{2-14}
    & 50 & 27 & 4 & 38 & 43 & 2 & 24 & 63 & 2 & 22 & 1 &  23.29 & 62\\
    \hline

    \multirow{2}{3em}{ERD} & 40 & 27 & 3 & 31 & 67 & 2 & 23 & 109 & 2 & 21 & 1 & 22.29  & 109\\   \cline{2-14}
                           & 50 & 27 & 5 & 39 & 41 & 2 & 24 & 69 & 2 & 17 & 5 & 23.46 & 65\\   \hline

    \multirow{2}{3em}{BAR} & 40 & 27 & 3 & 17 & 71 & 2 & 15 & 55 & 2 & 13 & 1 & 14.29  & 56\\     \cline{2-14}
     & 50 & 27 & 3 & 28 & 62 & 2 & 24 & 71 & 2 & 20 & 2 & 22.46 & 71\\     \hline

     \multirow{2}{3em}{IS} & 36 & 27 & 3 & 12 & 38 & 2 & 10 & 54 & 2 & 10 & 0 & 10  & 55\\     \cline{2-14}
     & 49 & 27 & 3 & 19 & 42 & 2 & 14 & 54 & 2 & 14 & 0 & 14 & 54\\     \hline

     \multirow{2}{3em}{XY} & 36 & 27 & 5 & 61 & 42 & 2 & 23 & 112 & 2 & 17 & 3 & 20.88  & 111\\     \cline{2-14}
     & 50 & 27 & \multicolumn{3}{c||}{No Solution} & 2 & 30 & 111 & 2 & 26 & 2 & 28.58 & 108\\     \hline

     \multirow{2}{3em}{HS} & 36 & 27 & 4 & 62 & 60 & 2 & 23 & 165 & 2 & 19 & 2 & 21.58  & 167\\     \cline{2-14}
     & 49 & 27 & \multicolumn{3}{c||}{No Solution} & 2 & 30 & 160 & 2 & 26 & 3 & 29.88 & 161\\     \hline

     \multirow{2}{3em}{IS-n} & 36 & 27 & 2 & 18 & 81 & 3 & 16 & 127 & 2 & 16 & 0 & 16 & 127\\     \cline{2-14}
     & 50 & 27 & \multicolumn{3}{c||}{No Solution} & 2 & 24 & 71 & 2 & 20 & 2 & 22.46 & 71\\     \hline

     \multirow{2}{3em}{XY-n} & 40 & 27 & \multicolumn{3}{c||}{No Solution} & 2 & 38 & 259 & 2 & 34 & 2 & 36.58 & 255\\     \cline{2-14}
     & 50 & 27 & \multicolumn{3}{c||}{No Solution} & 2 & 84 & 264 & 2 & 73 & 4 & 78.17 & 264\\     \hline

     \multirow{2}{3em}{HS-n} & 40 & 27 & \multicolumn{3}{c||}{No Solution} & 2 & 15 & 55 & 2 & 13 & 1 & 14.29  & 56\\     \cline{2-14}
     & 50 & 27 & \multicolumn{3}{c||}{No Solution} & 2 & 24 & 71 & 2 & 20 & 2 & 22.46 & 71\\     \hline

     \multirow{2}{3em}{VQE} & 42 & 27 & 2 & 1 & 26 & 2 & 1 & 26 & 2 & 1 & 0 & 1  & 26\\     \cline{2-14}
     & 50 & 27 & 2 & 1 & 26 & 2 & 1 & 26 & 2 & 1 & 0 & 1  & 25\\     \hline
    
    \end{tabular}
  \label{table:Wire_gate_cut_results}
\end{table*}

\subsection{Wire- and Gate- Cutting Evaluation}
Table \ref{table:Wire_gate_cut_results} compares the cutting solutions when applying cutting on the benchmarks that compute expectation values. We compare two choices for our scheme: one is to choose W-Cut only, while the other allows both W-Cut and G-Cut.

For comparison purposes, for a solution ($k_1$, $k_2$), where $k_1$ and $k_2$ are the W-Cut and G-Cut numbers, respectively, its overhead 4$^{k_1}$6$^{k_2}$ is converted to  4$^{k_3}$ with $k_3$ being the effective W-Cut number reported in the table. On average, \OurScheme{} (W-Cut only) and \OurScheme{} (both) achieve 41\% and 44\% reductions in the number of cuts. Exploiting G-Cut further reduces post-processing overhead.
For example, for ERD-50, \OurScheme{} (both) has an \#EffCuts of 22.46. While being a small reduction over 24 from QRCC(W-Cut only), it corresponds to an 8.45$\times$ reduction in post-processing overhead. 


\subsection{Real Machine Evaluation}
The cutting solutions that adopt either W-cut or G-cut have already been independently verified in \cite{cutqc} and \cite{Mitarai_2021}, respectively. We next verify the solution by adopting both and highlight our strategy for efficient post-processing.

We verify our approach by testing benchmark REG($m$=2) on an IBM 7-qubit LAGOS quantum computer. 
The computer has 1.7 physical connections per qubit. When we ran the experiments, it had median 
error rates of 8.25e$^{-3}$ for CNOT gates and 2.6e$^{-4}$ for single-qubit, $\sqrt{x}$ gates, respectively.
 
We choose $N$=7 and $D$=4, i.e., the original quantum circuit has seven qubits, and \OurScheme{} partitions it into smaller subcircuits so that each subcircuit can run on a 4-qubit quantum computer. 
Table \ref{table:real_machine_results} compares four execution modes. 
\begin{itemize}[leftmargin=*]
\item {\em State Vector Simulation}: The result from the state vector simulation computes the ground truth for the comparison of the results from different schemes.  
\item {\em Shot-based Simulation:} In this mode, we run a shot-based state-vector simulation, which uses the state-vector probability to introduce a random bit-string output every shot for simulating an ideal device. We report the average from 10 circuit runs, with each run having 16,384 shots.
\item {\em Device Execution (7-qubit)}: We ran the 7-qubit original circuit on the real quantum computer with 16,384 shots for each run. We run 10 times and report the average.
\item {\em \OurScheme{}}: \OurScheme{} partitions the original circuit into two subcircuits with one gate cut and one wire cut. The subcircuits are run with different measurement and initialization instances, resulting in a total of 42 instances. Each instance runs just one time (with 16,384 shots) using four physical qubits.
The results from subcircuits are then combined to compute the result for the original circuit.
\end{itemize}

\begin{table}[tbp]
   \centering
     \footnotesize
   \caption{Comparison between 7-qubit device execution and \OurScheme{} (4-qubit device execution + post-processing).}
    \begin{tabular}{||l|c|c||}
    \hline
    Execution Mode & Results & Accuracy \\  \hline \hline
    State Vector simulation & -0.0349  & 100$\%$\\
    Shot-based Simulation   & -0.0323 & 92$\%$\\
    Device Execution (7-qubit) & -0.0078 & 22.3$\%$\\
    \OurScheme{}-B & -0.0355 & 98.3\%\\     \hline
    \end{tabular}
    \vspace{-0.1in}
   \label{table:real_machine_results}
\end{table}

From the table, \OurScheme{} achieves better accuracy than that of 7-qubit device execution.
This is because (1) The original circuit has 16 two-qubit CNOT gates (and 9 of them were introduced from SWAP operation) while each of our subcircuits contains 3 CNOT gates. This results in better computation fidelity for the subcircuit execution. (2) The subcircuits have fewer qubits and short execution depths. 
Due to noisy qubits, the expectation value result from quantum device execution shows low accuracy; similar results were also observed in recent studies \cite{hammer}.

In addition, \OurScheme{} achieves better accuracy than that of shot-based simulation. The state vector of the shot-based simulation is 8$\times$ that of the 4-qubit device execution, which tends to introduce more randomness in the output probability distribution than that of the real execution.

\subsection{Studying $\delta$ Parameter}

\begin{figure}[tbp]
    \centering
    \includegraphics[width=3.2in]{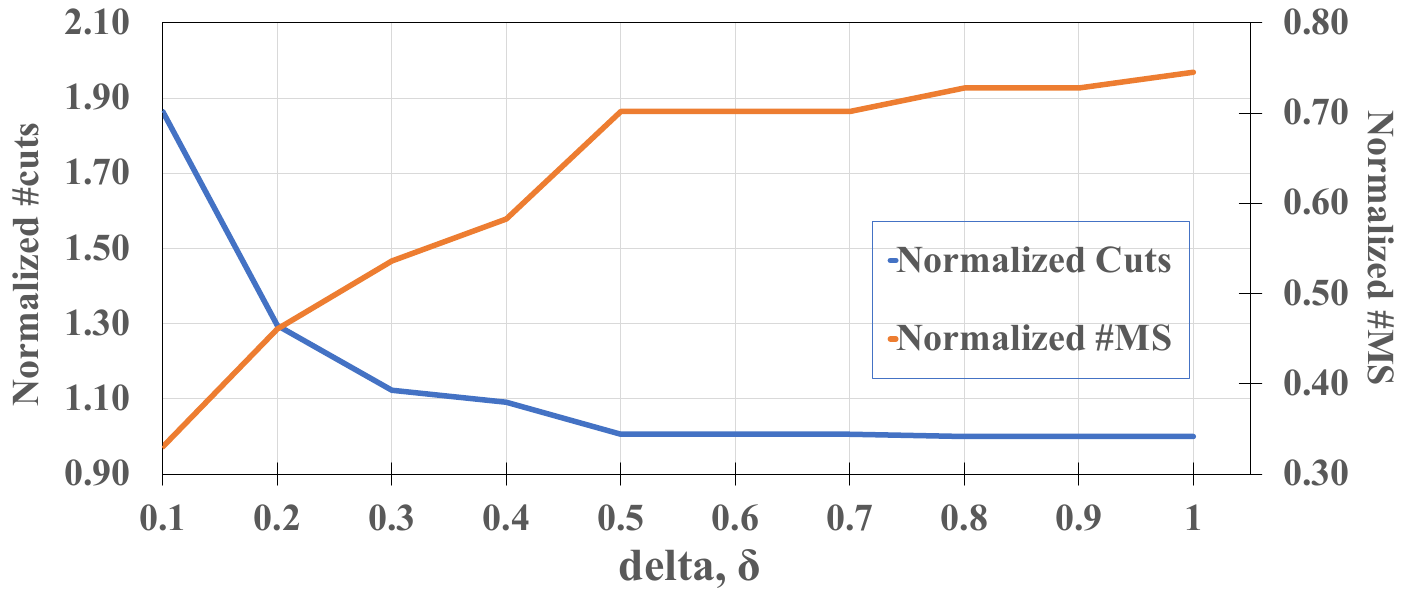}
    \caption{Correlating varying $\delta$ values with \#cuts and \#MS.
    The left y-axis represents the \#cuts, normalized to that when $\delta$=1. 
    The right y-axis represents the \#MS, normalized to the size of the original circuits.}
    \label{fig:delta_sweep}
    \vspace{-0.15in}
\end{figure}

In Equation (18), assigning different $\delta$ values changes the priority on post-processing overhead and computation fidelity. 
We next study the impact on the effective cut numbers (i.e., \#cuts) and the largest \#MS in subcircuit with varying $\delta$ values.
The post-processing overhead increases with larger \#cuts values, and the computation fidelity increases with smaller \#MS values.
Figure \ref{fig:delta_sweep} reports the average for the benchmarks with both W-Cut and G-Cut. The $\delta$ value varies from 0.1 to 1.0 as the post-processing overhead is the main design goal and thus cannot be completely ignored.

From the figure, \#cuts decrease and \#MS increases when the $\delta$ value increases.
This is because giving higher priority to post-processing overhead, i.e., assigning a large $\delta$ value, minimizes \#cuts.
Given higher priority to computation fidelity, i.e., assigning a smaller $\delta$ value, minimizes \#MS but hurts \#cuts significantly.
The figure also reveals that \#cuts stabilize when $\delta$>0.5, while the impact on \#MS is significant.
In the paper, we choose $\delta$=0.7 for QRCC-B, which exhibits negligible impact on \#cuts but a large improvement on \#MS.
For a strategy that chooses a smaller $\delta$ value, e.g., $\delta$=0.2, it increases \#cuts by 30\% on average.
We observed that, for some benchmarks, it may find a solution that has a higher post-processing overhead than that from CutQC. 
Meanwhile, this choice leads to significantly improved computation fidelity, on average 52\% improvement of \#MS.

\subsection{Time Comparison}
We next compare the time required to find the cutting solutions using \OurScheme{} and CutQC. 
For a fair comparison, we assume we know $k$, the number of subcircuits of the solution for each setting. 
This is because \OurScheme{} and CutQC work slightly differently. For \OurScheme{}, the user specifies a range
[$C_{min}$,$C_{max}$] and the subcircuit number of the found solution $k$ is guaranteed to be within the range. For CutQC, the user needs to specify the exact $k$ such that CutQC searches for the best solution with $k$ subcircuits. For the latter, the user needs to manually increment $k$ if a smaller $k$ value results in $no$-$solution$.
In the experiment, assuming we know $k$, we set $C_{min}$=$k$=$C_{max}$ for \OurScheme{} and start with $k$ for CutQC.

\begin{table}[h!]
  \centering
  \small
  \caption{The searching time comparison of the ILP model in QRCC and the MIP model in CutQC.}
  \vspace{-0.08in}
  \begin{tabular}{||m{2em} |m{2em} |m{2em} |m{2em} |m{2em} |m{3em}||}
    \hline
    \multicolumn{3}{||c|}{Benchmark} & \multicolumn{1}{l|}{CutQC} & \multicolumn{1}{l|}{\OurScheme{}} & \multirow{2}{3em}{Improv.} \\
    \cline{1-3}
    Name &  \multicolumn{1}{|l|}{Circuit}& \multicolumn{1}{l|}{Device} & time & time & \\
    \hline
    \hline
    \multirow{4}{3em}{SPM} & 15 & 7 & 0.80 & 0.63 & 21\%\\
     & 20 & 7 & 11.7 & 6.21 & 47\%\\
     & 30 & 16 & 7.05 & 0.54 & 92\%\\
     & 42 & 16 & 65.16 & 18.1 & 72\%\\
    \hline
    \multirow{3}{3em}{QFT} & 15 & 9 & 1800 & 1.42 & 100\%\\
    & 30 & 24 & 1800 & 19.28 & 100\%\\
    & 30 & 27 & 1800 & 1.92 & 100\%\\
    \hline
    \multirow{4}{3em}{ADD} & 16 & 7 & 31.1 & 4.16 & 87\%\\
    & 22 & 7 & 148.5 & 19.70 & 87\%\\
    & 30 & 16 & 4.72 & 0.75 & 84\%\\
    & 40 & 16 & 14.0 & 13.1 & 6\%\\
    \hline
    \multirow{4}{3em}{AQFT} & 15 & 7 & 18.0 & 18.0 & 0\%\\
    & 20 & 7 & 1800 & 1800 & 0\%\\
    & 30 & 16 & 33.9 & 26.3 & 22\%\\
    & 40 & 27 & 1565 & 1297 & 17\%\\
    \hline
    \end{tabular}
    \vspace{-0.08in}
  \label{table:Wire_cut_timing_results}
\end{table}

Table \ref{table:Wire_cut_timing_results} summarizes the wall clock time to find the solutions in Table 1.
From the table, \OurScheme{} runs much faster than CutQC for most cases. On average, \OurScheme{} is 58\% faster than CutQC, for cases where CutQC can find a solution. The main reason is that \OurScheme{} builds a linear model while CutQC adopts a non-linear model. The quadratic constraints in CutQC significantly slow down the search performance. In addition, without qubit reuse, CutQC introduces one extra qubit (i.e.,{\em initialization qubit}) after each cut, which increases the number of qubits in the subcircuit and makes it difficult to find a valid cutting solution.

\subsection{Scalability}

For cutting-based approaches such as QRCC and CutQC, the classical post-processing overhead dominates the overall overhead for handling large and complicated quantum circuits. 
In this section, we study the scalability of such approaches with scaled problem sizes.

\subsubsection{Scalability vs \#cuts}
\begin{figure}[tbp]
    \centering
    \includegraphics[width=3in]{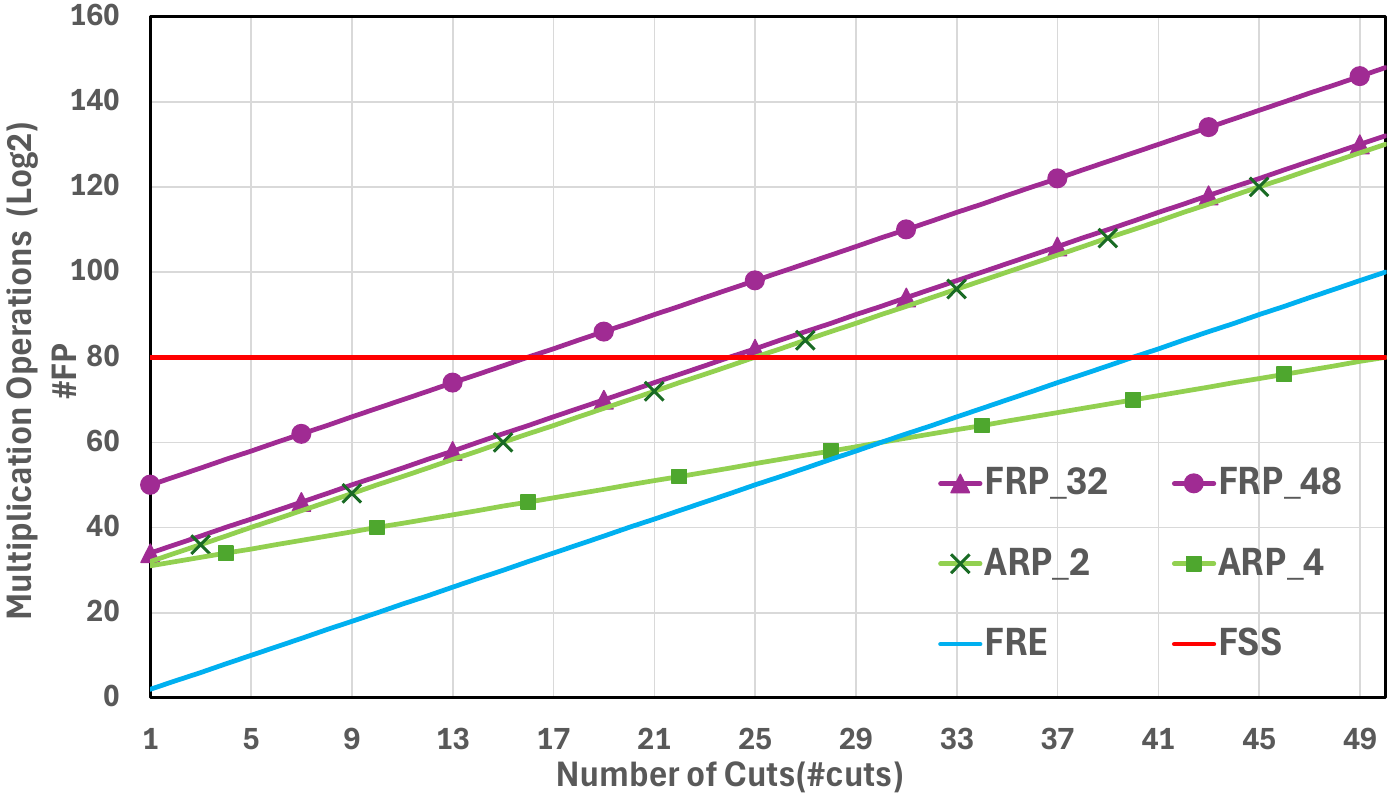}
    \caption{
    Comparison of the computation overhead with different reconstruction schemes: 
    i) FRP (purple curves) -- the hybrid full state reconstruction;
    ii) ARP (green curves) -- the hybrid approximate reconstruction;
    iii) FSS (red curve) -- the full-state simulation threshold; 
    iv) FRE (blue curve) -- the expectation value of the original circuit.
    The Y-axis indicates the log scale of the post-processing overhead in terms of \#FP operations.}
    \label{fig:post_fullstate}
    \vspace{-0.2in}
\end{figure}

As discussed in Section 2, the classical post-processing overhead increases exponentially with increasing numbers of effective circuit cuts. We next compare the computation overhead using different schemes to reconstruct the results of the original circuit. We evaluate the computation overhead using the number of floating-point number operations (\#FP) and ignore the memory requirement. We summarize the result in Figure \ref{fig:post_fullstate}. The X-axis indicates the number of circuit cuts (\#cuts) and the y-axis indicates the log scale of required \#FP for post-processing.

\begin{itemize}[leftmargin=*]
\item {\bf FSS}: the full-state simulation of a dense 34-qubit 1000-gate quantum circuit. It requires about 1e24 \#FP, shown as the Horizontal red curve in the figure.
Simulating such a circuit sequentially at the gate level may take several hours on CPU \cite{flops_requirement,Google_Quantum_AI,AWS}.
This is set as a threshold such that a reconstruction process is considered {\em too expensive} if its post-processing overhead exceeds this threshold. This threshold is set for illustration purposes and thus can be adjusted according to different settings.

\item {\bf FRP}: the hybrid full-state reconstruction for the probability vector of the original circuit. FRP assumes that we reconstruct two subcircuits cut from an $N$-qubit original circuit with $\#cuts$ cuts (X-axis). For simplicity, we assume that the original qubits are evenly distributed among two subcircuits.

\item {\bf FRE}: The reconstruction of FRE is for the expectation value of the original circuit. Shown as the blue curve in figure \ref{fig:post_fullstate}, this scheme has a similar assumption as above. 

\item {\bf ARP-2}: the hybrid approximate reconstruction for the probability vector of the original circuit. ARP-2 assumes the reconstruction from two subcircuits and the original qubits are evenly distributed among them.\\
ARP-2 differs from FRP in that it adopts an approximation strategy as follows. 
A full-state reconstruction strategy, such as FRP, faces a big challenge for circuits with large numbers of qubits. For example, saving the full probability vector of a 50-qubit quantum circuit demands O(2$^{50}$) or PB-scale memory space, which is prohibitive for most small- to medium-scale servers. An alternative approximate reconstruction \cite{scaleqc} is to shrink the original vector space 2$^{50}$ to a small vector space, e.g., 2$^{30}$. ARP-2 exploits the approximate reconstruction for all N>30.

\item {\bf ARP-4}: This is similar as ARP-2 but the reconstruction is conducted on four subcircuits. For simplicity, we horizontally partition the original circuit into four subcircuits (i.e., S1, S2, S3, and S4) such that the original qubits are evenly partitioned to the four subcircuits; and the cuts are evenly partitioned for cutting S1/S2, S2/S3, and S3/S4. A wire cut for S1/S2 indicates its measurement is in S1 and its initialization is in S2.
\end{itemize}

From the figure, the FRP\_48 curve has the highest reconstruction cost, as a function of the number of $\#cuts$, because of the 2$^{48}$ state space of the output vector. FRP reproduces the full state vector and thus demands O(2$^{N + 2*\#cuts}$) \#FP. FRE has much lower overheads, O(2$^{2*\#cuts}$), due to computing one expectation value instead of long probability vectors. This can be observed by the vertical distances between the purple lines (FRP\_32 and FRP\_48)  to the blue curve (FRE). FRE independently computes the expectation value of each subcircuit instance, and multiples these expectation values together based on equation \ref{equation:cutqc} and \ref{equation: gate_equation}. 
As such, only scalar multiplication is required in FRE, and the number of scalar multiplications is independent of the number of qubits and is only affected by the number of $\#cuts$. 

As a comparison, when N=48, FRE can tolerate 40 \#cuts, while FRP\_48 can only tolerate 16 \#cuts before hitting the post-processing overhead threshold.

Furthermore, by exploiting approximate reconstruction, ARP-2 and ARP-4 can tolerate a larger number of \#cuts, e.g., 25 and 50 \#cuts respectively, before hitting the FSS threshold, as shown in the figure. This is because their overhead is qubit-independent when N>30. No matter the size of the circuit, only 2$^{30}$ states of the original circuit are reproduced. One can also observe that when the original circuit is divided into more subcircuits, e.g., four subcircuits in ARP-4 instead of two in ARP-2, the reconstruction overhead decreases. This is because the overhead is dependent on the \#cuts required to combine each two subcircuit pairs. Each combination of the subcircuit instances (e.g., S1/S2, S2/S3, and S3/S4) is independent of each other. For example, S1/S2 and S3/S4 can first be combined independently, as the cuts between S1/S2 do not have any effect on the combination of S3/S4 and vice-versa. 
This allows a divide-and-conquer strategy for the recombination of the original output, as the overhead only depends on the largest number of cuts among all the subcircuit pairs, not the total number of cuts aggregated across all subcircuit pairs. Consequently, the overhead increases at a slower pace than that of the total \#cuts, validating that the use of recursive circuit-cutting improves the scalability of our proposed framework. However, we also want to mention that, while the increased number of subcircuits can have lower overhead in reconstruction, the complexity of circuit cutting increases due to the increased search space.

From the figure, it is clear that the post-processing overhead is dominated by the number of cuts (\#cuts).
Therefore, the reduction of \#cuts from QRCC over CutQC can effectively mitigate the corresponding post-processing computation time.

For example, for REG (N=40, D=27) in Table \ref{table:Wire_gate_cut_results}, the effective \#cuts are reduced from 21 in CutQC to 16.3 in QRCC-C. This corresponds to $\mathcal{O}$(4$^{21 - 16.29}$) reduction in computation overhead, or a 685$\times$ speedup of post-processing time, without considering the memory requirement.

\begin{figure}[tbp]
    \centering
    \includegraphics[width=3.2in]{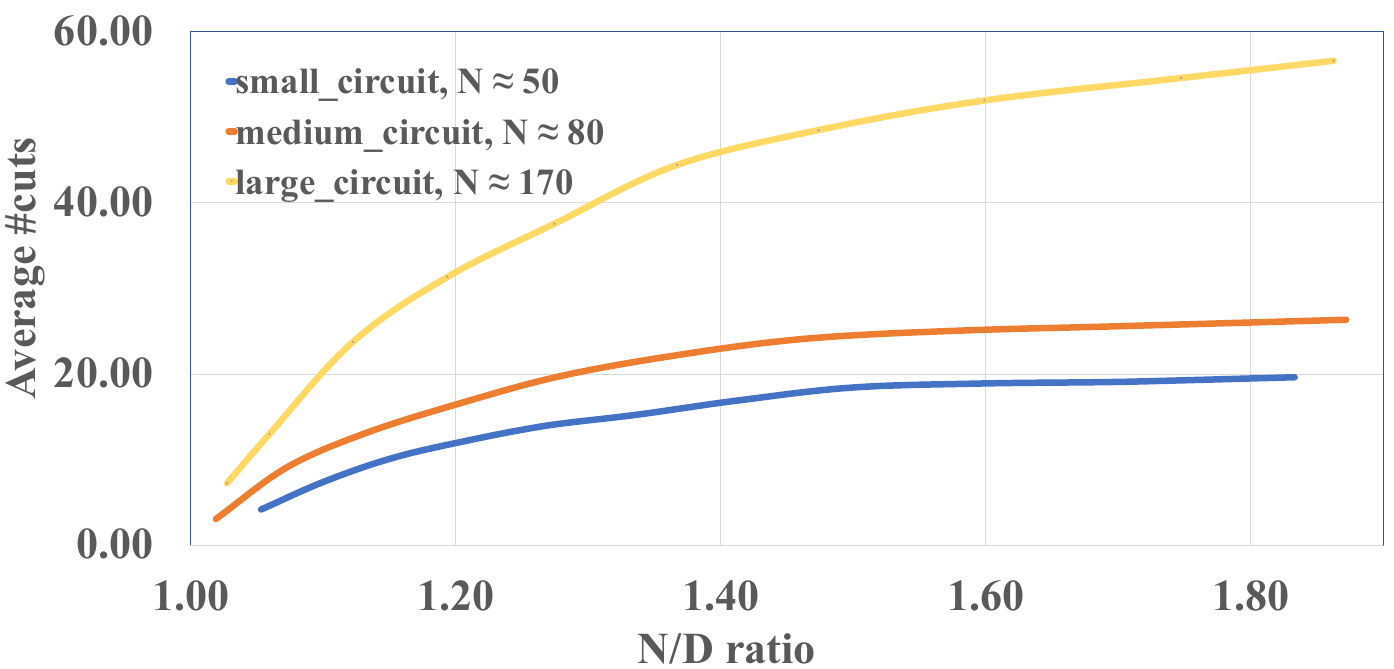}
    \caption{The \#cuts values increase with larger $N$/$D$ ratios. The $N$ values are 50, 80, and 170 for small, medium, and large circuits, respectively.}
    \label{fig:NandD_sweep}
    \vspace{-0.15in}
\end{figure}

\subsubsection{Scalability vs N/D Ratio}
We next correlate the number of circuit cuts \#cuts to the circuit size $N$ and the device size $D$. Intuitively, the cutting problem becomes more challenging when we have larger $N$ values and smaller $D$ values. 
Figure \ref{fig:NandD_sweep} reports the impact on \#cuts with different $N$/$D$ ratios with the results averaged on all benchmarks from table \ref{table:Wire_gate_cut_results}. We choose $N$= 50, 80, and 170 for small, medium, and large circuits, respectively.

From the figure, we observe that \#cuts increase with larger $N$/$D$ ratios. For small and medium circuits, the increase is moderate as there exists many qubit reuse opportunities.  For large circuits, the increase is at a faster pace due to more two-qubit gates in the circuits. 

\subsubsection{Scalability vs Circuit Connectivity}
To further study the impact of increased two-qubit gates, we fix $N$ and $D$ values for a subset of large circuits (in Table \ref{table:Wire_gate_cut_results}) whose circuit complexity can be adjusted with a meta-parameter, and summarize the required number of circuit cuts in Table~\ref{table:scalability}. 
As discussed above, increasing the N/D ratio from 200/150 to 300/200 for REG (m=3) leads to more cuts for both schemes.
We also observe that, by adjusting the meta-parameter $m$ (from m=3 to m=4), the circuits contain more two-qubit gates and thus demand around double the amount of cuts.
For more complex circuits, e.g., choosing $N$=300, $D$=200, and $p$=0.02 for ERD, our model still scales well and finds a solution. However, the solution contains large numbers of wire cuts and gate cuts, indicating that the bottleneck has shifted to the post-processing overhead, i.e., O(4$^{52}$6$^{104}$).

\begin{table}[tbp]
  \centering
    \footnotesize
  \caption{The scalability correlates to circuit size and complexity.}
  \vspace{-0.05in}
  \begin{tabular}{||ccc|cc|c||}
    \hline
    \multicolumn{3}{||c|}{Benchmark} & \multicolumn{2}{c|}{\OurScheme{}} & \multicolumn{1}{c||}{CutQC} \\
    \hline
    name &  \multicolumn{1}{l}{N}& \multicolumn{1}{l|}{D} & \#W-Cuts & \#G-Cuts & \#W-Cuts \\ \hline
    \hline
    
    REG (m=3) & 200 & 150 & 19 & 0 & 21 \\
    
    REG (m=3) & 300 & 200 & 31 & 3 & 36 \\  \hline

    REG (m=4) & 200 & 150 & 36 & 3 & 49 \\
    
    REG (m=4) & 300 & 200 & 61 & 6 & 75 \\  \hline
    
    BAR (m=4) & 200 & 150 & 74 & 3 & No Solution\\
    
    BAR (m=2) & 300 & 200 & 55 & 1 & 60\\  \hline
    
    ERD ($p$=0.05) & 200 & 150 & 96 & 2 & No Solution\\
    
    ERD ($p$=0.02) & 300 & 200 & 52 & 104 & No Solution\\

    \hline
    \end{tabular}
    \vspace{-0.15in}
  \label{table:scalability}
\end{table}

\subsection{Qubit Reuse in Cutting}
Based on the observation that \OurScheme{} exploits qubit reuse to find better cutting solutions than those of CutQC,
it becomes interesting to investigate if naively combining CutQC and qubit reuse can achieve similar results.

To partition a $N$-qubit original circuit into smaller subcircuits that can run on $D$-qubit quantum devices, we have two simple approaches to combine CutQC and qubit reuse.
\begin{itemize}[leftmargin=*]
    \item[(i)] For the first approach, we apply CutQC to partition the original circuit into small subcircuits that can each run on $N$-qubit devices, and then optimize each subcircuit using qubit reuse. Compared to \OurScheme{}, this is a sub-optimal approach because its first step often results in a cutting solution with more cuts than that of \OurScheme{}. Applying qubit reuse at the second step, even if it reduces the number of required qubits for each subcircuit, shall not help to reduce the post-processing overhead.
    \item[(ii)] Alternatively, we may partition the original circuit into subcircuits that can run on $X$-qubit devices, where $N$>$X$>$D$, assuming we can reduce $X$ to $D$ by applying qubit reuse. Unfortunately, the assumption is not always true.
\end{itemize}

For example, we choose QFT with $N$=15 and $D$=7, \OurScheme{} finds a cutting solution that partitions the circuit into three subcircuits with 20 wire cuts. Given that CutQC cannot find a solution for $D$=7 or 8, we may choose other solutions and then apply qubit reuse. We try all different $N$>$X$>$D$ settings and summarize the results in Table \ref{table:cutqc+carqr}.

From the table, sequentially applying CutQC and qubit reuse cannot find a solution as good as the one from \OurScheme{}.
The closest one is the solution at X=9 when all subcircuits after the cut can run on 9-qubit devices. Applying qubit reuse enables them to run on 7-qubit devices. However, the number of cuts is more than twice the number of our solution, i.e., 44 vs 20. For all other settings, the required qubits for the subcircuits can be reduced after reuse, but the subcircuits still cannot run on 7-qubit quantum computers. 

\begin{table}[tbp]
  \centering
    \footnotesize
  \caption{Applying CutQC and qubit reuse sequentially produces sub-optimal results. }
  \begin{tabular}{||c|c|c|c|c||}
    \hline
    \multirow{2}{5em}{Device size} & \multicolumn{3}{c|}{CutQC} & \multicolumn{1}{c||}{ + CaQR} \\
    \cline{2-5}
    &\#SC & \#cuts & width & width\\  \hline \hline
    
    9  & 9 & 44 & 9  & 7 \\
    
    10 & 4 & 24 & 10 & 8 \\
    
    11 & 4 & 20 & 11 & 10\\
    
    12 & 4 & 20 & 12 & 10 \\
    
    13 & 4 & 20 & 12 & 10 \\
    
    14 & 4 & 20 & 12 & 10 \\ \hline
    
    \end{tabular}
  \label{table:cutqc+carqr}
\end{table}

\section{Related Work}
The recent studies on circuit-cutting focus mainly on lowering the reconstruction overhead. 
Lowe {\em et al.}\cite{random_cutting} proposed to reduce the overhead of wire cutting using randomized probabilistic measurements.
Piveteau {\em et al.}\cite{LOCC} proposed to reduce the overhead of gate cutting, using classical two-way communication and shared bell pair between subcircuits. 
These works assume that the input is a pre-cut circuit and thus are orthogonal to our work.

Xie {\em et al.} proposed a compiler framework for distributed quantum computing \cite{Auto}. 
Smith {\em et al.} exploited circuit-cutting and Clifford gate simulation to enhance the reach of classical quantum circuit simulation and the simulation time~\cite{clifford_cutting}.   

\section{Conclusion}
In this paper, we propose \OurScheme{} for evaluating large quantum circuits on small quantum computers. 
\OurScheme{} integrates wire cutting and qubit reuse in one framework to find good cutting solutions for quantum circuits that compute probability vectors, and in addition, with gate cutting for circuits that compute expectation values.
We formulate the problem as an ILP model to find the cutting solutions efficiently. 

\section{Acknowledgments}
This work is supported by the National Science Foundation under award numbers \#2334628, \#2011146, \#2154973, and \#2312157, and Pittsburgh Quantum Institute under award number \#007913. We extend our gratitude to the ASPLOS reviewers for their valuable insights and feedback.

\bibliographystyle{plain}
\bibliography{references}

\end{document}